\documentclass[12pt]{article}
\usepackage{latexsym,amssymb}
\usepackage[american]{babel}
\makeatletter
\renewcommand{\@evenhead}{\raisebox{0pt}[\headheight][0pt]{\vbox{\hbox
to \textwidth{\thepage\hfil\strut\textsc{\leftmark}}\hrule}}}
\renewcommand{\@oddhead}{\raisebox{0pt}[\headheight][0pt]{\vbox{\hbox
to \textwidth{\textsc{\rightmark}\hfil\strut\thepage}\hrule}}}
\makeatother

\def\sq{\Square}
\def\Square{\mathchoice\sqr6\sqr5\sqr4\sqr3}
\def\sqr#1{{\mathop{\mkern0.5\thinmuskip\vbox{\hrule\hbox{\vrule\hskip#1pt
\vrule height#1pt width 0pt \vrule}\hrule}\mkern0.5\thinmuskip}}}

\def\h#1{{\cal #1}}

\def\b{\beta}
\def\g{\gamma}
\def\d{\delta}

\def\l{\lambda}
\def\m{\mu}

\def\om{\omega}
\def\na{{\nabla}}
\def\N{{\nabla}}
\let\d=\delta

\let\l=\lambda

\let\m=\mu

\let\P=\Pi

\let\x=\xi

\let\b=\beta

\let\g=\gamma
\let\N=\nabla
\def\na{\nabla}

\def\II{{\mathbb I}}
\def\RR{{\mathbb R}}
\def\CC{{\mathbb C}}

\def\dist{{\rm \,dist\,}}

\def\tr{{\rm \,tr\,}}
\def\Tr{{\rm \,Tr\,}}

\def\det{{\rm \,det\,}}

\def\Det{{\rm \,Det\,}}
\def\End{{\rm \,End\,}}
\def\erf{{\rm \,erf\,}}
\def\erfc{{\rm \,erfc\,}}

\def\sinh{{\rm \,sinh\,}}

\def\coth{{\rm \,coth\,}}
\def\sign{{\rm \,sign\,}}

\def\log{{\rm \,log\,}}

\def\vol{{\rm \,vol\,}}

\def\tr{{\rm \,tr\,}}
\def\Tr{{\rm \,Tr\,}}

\def\det{{\rm \,det\,}}

\def\Det{{\rm \,Det\,}}
\def\Aut{{\rm \,Aut\,}}
\def\End{{\rm \,End\,}}

\def\sinh{{\rm \,sinh\,}}

\def\coth{{\rm \,coth\,}}

\def\log{{\rm \,log\,}}

\def\vol{{\rm \,vol\,}}
\def\R{{\cal R}}

\def\II{{\bf I}}
\def\RR{{\bf R}}
\def\CC{{\bf C}}

\def\tr{{\rm tr\,}}
\def\Tr{{\rm Tr\,}}

\def\det{{\rm det\,}}

\def\Det{{\rm Det\,}}
\def\End{{\rm End\,}}

\def\sinh{{\rm sinh\,}}

\def\coth{{\rm coth\,}}

\def\log{{\rm log\,}}

\def\vol{{\rm vol\,}}

\def\be{\begin{equation}}
\def\ee{\end{equation}}
\def\bea{\begin{eqnarray}}
\def\eea{\end{eqnarray}}
\def\bes{\begin{displaymath}}
\def\ees{\end{displaymath}}
\def\bmp{\begin{minipage}}
\def\emp{\end{minipage}}

\def\sideremark#1{\ifvmode\leavevmode\fi\vadjust{\vbox to0pt{\vss
 \hbox to 0pt{\hskip\hsize\hskip1em
 \vbox{\hsize2cm\tiny\raggedright\pretolerance10000
 \noindent #1\hfill}\hss}\vbox to8pt{\vfil}\vss}}}%
\begin{document} 
 
\begin{titlepage} 
\null

\hspace*{6truecm}{\hrulefill}\par\vskip-4truemm\par 
\hspace*{6truecm}{\hrulefill}\par\vskip5mm\par 
\hspace*{6truecm}{{\large\sc New Mexico Tech {\rm (July, 2001)}}}\vskip4mm\par 
\hspace*{6truecm}{\hrulefill}\par\vskip-4truemm\par 
\hspace*{6truecm}{\hrulefill} 
\par 
\bigskip 
\bigskip
\par

\begin{tabular}{p{5.5truecm}p{7truecm}}
& {To appear in: Proceedings of the International Conference
``{\sf Quantum Gravity and Spectral Geometry}'', July 2--7, 2001, Naples,
Italy}\\
\end{tabular}

\vfill 
\centerline{\LARGE\bf Heat Kernel Approach} 
\medskip 
\centerline{\LARGE\bf in Quantum Field Theory} 
\bigskip 
\bigskip 
\centerline{\Large\bf Ivan G. Avramidi} 
\bigskip 
\centerline{\it Department of Mathematics} 
\centerline{\it New Mexico Institute of Mining and Technology} 
\centerline{\it Socorro, NM 87801, USA} 
\centerline{\it E-mail: iavramid@nmt.edu} 
\bigskip 
\medskip 
\vfill 
 
{\narrower\par 
We give a short overview of the effective action approach in quantum 
field theory and quantum gravity and describe various methods for 
calculation of the asymptotic expansion of the heat kernel for second-order 
elliptic partial differential operators acting on sections of vector bundles 
over a compact Riemannian manifold. We consider both Laplace type operators and 
non-Laplace type operators on manifolds without boundary as well as Laplace 
type operators on manifolds with boundary with oblique and non-smooth boundary 
conditions.
\par} 
\vfill 
\end{titlepage} 


\markboth{\sf Ivan Avramidi: Heat Kernel Approach in Quantum Field Theory}
{\sf Ivan Avramidi: Heat Kernel Approach in Quantum Field Theory} 
 
\section{Effective Action in Gauge Field Theories and
Quantum Gravity}



In this lecture we briefly describe the standard formal construction of the
generating  functional and the effective action in gauge theories following
the covariant  spacetime approach to quantum field theory developed mainly by
DeWitt \cite{dewitt84}.  The basic object of any physical theory is the
spacetime $M$, which is assumed  to be a  $m$-dimensional manifold with the
topological structure of a cylinder  
$$
M =I\times\Sigma,
$$
where $I$ is an open
interval of the  real line (or the whole  real line) and $\Sigma$ is some
$(m-1)$-dimensional manifold. The spacetime
manifold is here assumed to be globally hyperbolic and equipped with a
(pseudo)-Riemannian metric $g$ of signature $(- + \cdots +)$; thus,
a foliation of spacetime exists into spacelike sections identical to
$\Sigma$. Usually one also assumes the existence of a spin structure on $M$. 
A point $x=(x^\mu)$ in
the spacetime is described locally by the time  $x^0$ and the space
coordinates $(x^1, \dots, x^{m-1})$. We label the spacetime  coordinates by
Greek indices, which run from $0$ to  $m-1$, and sum over  repeated indices.  

Let us consider a vector bundle $V$ over the spacetime $M$ each fiber of which
is isomorphic to a vector space, on which the 
spin group ${\rm Spin}(1,m-1)$, i.e. the
covering  group of Lorentz group,  acts. The vector bundle $V$ can also have
an additional structure on which a  gauge group acts. The sections of the
vector bundle $V$ are called fields. The  tensor fields describe the particles
with integer spin (bosons) while the spin-tensor fields  describe particles
with half-odd spin (fermions).  Although the whole scheme can be developed
for the superfields (a combination  of boson and fermion fields), we restrict
ourselves in  the present lecture to boson fields.  A field $\varphi$  is
represented locally by a set of functions 
$$
\varphi=(\varphi^A(x)),
$$
where
$A=1,\dots, \dim V$. Capital Latin indices will be used to label the local
components of the fields. We will also use the condensed DeWitt notation,
where the discrete index $A$ and the spacetime point $x$ are combined in one 
lower case Latin index 
$$
i \equiv (A, x).
$$
Then the components of a field
$\varphi$ are 
$$
(\varphi^{i}) \equiv (\varphi^{A}(x)).
$$
As usual, we will also
assume that a summation over repeated  lower case Latin indices, i.e. a
combined  summation-integration, is  performed, viz.  
$$
J\cdot\varphi\equiv
J_{i}\varphi^{i}\equiv \int_{M}d\vol(x)\, J_{A}(x)\varphi^{A}(x),
$$
where
$d\vol(x)=dx\,|g|^{1/2}$, $|g|=\det g_{\mu\nu}$, is the natural
Riemannian volume element defined by some background  metric $g$.
We will often omit the volume element when this does not cause any
misunderstanding.

In quantum field theory (QFT) the vector bundle $V$ is called  the
configuration space. One assumes that the configuration space is an 
infinite-dimensional manifold ${\cal M}$.  The fields $\varphi^i$ are the
coordinates on this manifold, the variational derivative
$\delta/\delta\varphi$ is a tangent vector, a small disturbance
$\delta\varphi$ is a one-form and so on. If $S(\varphi)$ is a scalar field on 
the configuration space, then its variational derivative $\delta
S/\delta\varphi$ is a one-form on ${\cal M}$ with the components that we
denote by 
$$
S_{,i}={\delta S\over \delta \varphi^i}.
$$
By using the functional 
differentiation one can define formally the concept of tangent space,  the
tangent vectors, Lie derivative, one-forms, metric, connection,
geodesics and so on. 

The dynamics of quantum field theory is determined by an action functional
$S(\varphi)$, which is a differentiable real-valued scalar field on the
configuration space.  The dynamical field configurations are defined as the 
field configurations  satisfying the stationary action principle, i.e. they
must  satisfy the dynamical equations of motion 
$$
{\delta S\over \delta\varphi}=0
$$ 
with given boundary conditions. The set of all dynamical  field configurations,
i.e. those that satisfy the dynamical equations  of motion, ${\cal M}_{0}$, is
a subspace of the configuration space  called the dynamical subspace. 

Quantum field theory is  basically a theory of small disturbances on the
dynamical subspace.  Most of the problems of standard QFT deal with 
scattering processes, which are described by the transition amplitudes between
some well defined initial  and final states in the remote past and the remote
future.  The collection of all these amplitudes is called  the scattering
matrix, or shortly $S$-matrix. 


Let us single out in the space-time two causally connected in-- 
and out-- regions, that lie in the past and in the future 
respectively relative to the region $\Omega$, which is of interest 
from the dynamical standpoint.  Let $\left.|{\rm in}\right>$ and 
$\left.|{\rm out}\right>$ be some initial and final states of the 
quantum field system in these regions. Let us consider the
 transition amplitude $\left<{\rm out|in}\right>$ and ask the 
question: how does this amplitude change 
under a variation of the interaction with a compact support in the region 
$\Omega$. The answer to this question gives the Schwinger
variational principle which states that
\bes
\delta\left<{\rm out}|{\rm in}\right> 
= {i\over\hbar}\left<{\rm out}|\right.\delta\hat S\left.|{\rm in}\right>,
\ees
where $\delta \hat S$ is the corresponding change of the action. 
This principle gives a very powerful tool to study the transition 
amplitudes. The Schwinger variational principle can be called 
the quantization postulate, because all the information 
about quantum fields can be derived from it.

Let us change the external conditions by adding a linear interaction 
with  some external classical sources $J$ in the dynamical 
region $\Omega$, i.e. 
$$
\delta S=J\cdot \varphi.
$$
The amplitude $\left<{\rm out}|{\rm in}\right>$ 
becomes a functional of the sources that we denote by $Z (J)$. 
By using the Schwinger variational principle one can obtain
the chronological mean values in terms of the derivatives 
of the functional $Z (J)$ 
\bea
\left<\hat\varphi^{i_{n}}\cdots \hat\varphi^{i_{1}}\right>
&\equiv&
 \frac{\left<{\rm out}|T(\hat\varphi^{i_{n}}
\cdots \hat\varphi^{i_{1}})|{\rm in}\right>}
{\left<{\rm out}|{\rm in}\right>}
\nonumber\\[10pt]
&=&
\left({\hbar\over i}\right)^n
Z^{-1}\frac{\delta^n}{\delta J_{i_{n}}\cdots \delta J_{i_{1}}}Z\,,
\nonumber
\eea
where $T$ denotes the operator of chronological ordering that 
orders the (non-commuting) operators in order of their time 
variables from right to left.
In other words the functional $Z (J)$ is the generating functional 
for chronological amplitudes. Let us now define another
 functional $W (J)$ by
$$
Z=\exp\left({i\over\hbar}W\right).
$$
We have obviously
\bes
\left<\hat\varphi^{i_{n}}\cdots \hat\varphi^{i_{1}}\right>
= \left( \frac{\hbar}{i}\right)^{n}
e^{-{i\over\hbar}W}\frac{\delta^{n}}{\delta J_{i_{n}}
\cdots \delta J_{i_{1}}}e^{{i\over\hbar}W}\,,
\ees
in particular,
\bea
\left<\hat\varphi^{i}\right> &=& \phi^{i},
\nonumber\\[10pt]
\left<\hat\varphi^{i}\hat\varphi^{k}\right> 
&=& \phi^{i}\phi^{k}+\frac{\hbar}{i}{\cal G}^{ik},
\nonumber
\eea
where 
\bes
\phi^{i}= 
\frac{\delta W}{\delta J_{i}},
\ees
and
\bes
{\cal G}^{i_{n}\ldots i_{1}}
=\frac{\delta^{n}}{\delta J_{i_{n}}\cdots \delta J_{i}}W. 
\ees
The functional $\phi$ is called the background or the mean field, ${\cal
G}^{ik}$  is called  the full propagator,  and ${\cal G}^{i_{i}\ldots
i_{n}}$, are called the  full connected Green functions or the 
correlation functions. Thus, whilst $Z(J)$ is the  generating functional for
chronological amplitudes the  functional $W(J)$ is the generating functional
for the connected  Green functions. The Green functions satisfy the boundary 
conditions which are determined by the states  $\left.|{\rm in}\right>$ and
$\left|{\rm out}\right>$.

The mean field itself is a functional of the sources, $\phi=\phi(J)$,   
the derivative of the mean field being the full propagator
$$
{\delta\phi^{i}\over \delta J_j}={\cal G}^{ji}.
$$
In the non-gauge theories the full propagator ${\cal G}^{ij}$, 
which plays the role of the (infinite-dimensional) Jacobian,  
is non-degenerate. Therefore, one can change variables and 
consider $\phi$ as independent variable and $J(\phi)$ (as well as all other
functionals) as the functional of $\phi$.

There are many different ways to show that there is a functional
$\Gamma(\phi)$ such that  
$$
\left<{\delta S(\hat\varphi)\over\delta\hat\varphi}\right>
={\delta\Gamma(\phi)\over\delta\phi}.
$$ 
This functional is defined by
\bes
\left<{\rm out}|{\rm in}\right>
=\exp\left\{{i\over \hbar}\left(\Gamma+J\cdot\phi\right)\right\},
\ees
or by the functional Legendre transform
\bes
\Gamma(\phi)=W(J(\phi))-J(\phi)\cdot{\delta\over\delta J}W(J(\phi)).
\ees
This is the most important object in quantum field theory.
It contains all the information
about quantized fields. First of all, the first variation of 
$\Gamma$ gives the effective equations for the background fields
\bes
{\delta \Gamma\over\delta\phi}=-J\,.
\ees
These equations replace the classical equations of motion 
and describe the effective dynamics of the background field 
with regard to all quantum corrections.
That is why $\Gamma$ is called the effective action. Furthermore, 
the second derivative of $\Gamma(\phi)$ determines the full propagator 
\bes
{\cal G}=\left(-{\delta^2\Gamma\over\delta\phi^2}\right)^{-1}\,.
\ees
The higher derivatives determine the so-called full vertex functions
$$
\Gamma_{,i_1\cdots i_k},
$$
which are also called strongly 
connected, or one-particle irreducible functions. In other words,
$\Gamma(\phi)$ is the generating functional  for the full vertex functions.
The full vertex functions  together with  the full propagator determine the
full connected Green functions and, therefore,  all  chronological amplitudes
and, hence, the $S$-matrix.
Thus, the entire quantum field theory is summed up in the functional
structure  of the effective action. 

One can obtain a very useful formal representation for the 
effective action in terms of functional integrals (called
also path integrals, or  Feynman integrals). A functional
integral is an integral over the (infinite-dimensional) 
configuration space ${\cal M}$.  Although a rigorous mathematical definition
for  the functional integrals is absent, they can be used in 
perturbation theory of QFT as an  effective tool, especially in gauge
theories, for manipulating the  whole series of perturbation theory. The point
is that in perturbation  theory one encounters only the functional
integrals of Gaussian type, which can be well defined effectively in terms
of the classical propagators and vertex functions. The Gaussian integrals do
not depend much on the dimension and, therefore, many formulas from the
finite-dimensional case, like the  Fourier transform, integration by parts,
delta-function, change  of variables etc. are valid in the
infinite-dimensional case as well. 
One has to note that the functional integrals are formally 
divergent --- if one tries to evaluate the integrals, one encounters 
meaningless divergent expressions. This difficulty 
can be overcome in the framework of the renormalization theory in 
so-called renormalizable field theories, but we will not discuss this
problem in the present lectures.

Integrating the Schwinger variational principle one can obtain the
following functional integral:
\bes
\left<{\rm out}|{\rm in}\right>=
\int\limits_{\cal M}{\cal D}\varphi
\exp\left\{{i\over\hbar}\left[S(\varphi)+J\cdot\varphi\right]\right\}.
\ees
Correspondigly, for the effective action one obtains a functional equation
$$
\exp\left\{{\frac{i}{\hbar}\Gamma(\phi)}\right\}
=
\int\limits_{\cal M}{\cal D}\varphi \exp\left\{{\frac{i}{\hbar}
\left[S(\varphi)-{\delta\Gamma(\phi)\over\delta\phi}\cdot
(\varphi-\phi)\right]}\right\}\,.
$$

The only way to get numbers from this formal 
expression is to take advantage of the semi-classical 
approximation within a formal  expansion in powers of Planck
constant $\hbar$: 
\bes
\Gamma=S+\sum\limits_{k=1}^\infty \hbar^{k-1}\Gamma_{(k)}.
\ees
Substituting this expansion in the functional equation for the 
effective action, shifting the integration variable in the 
functional integral 
$$
\varphi=\phi+\sqrt\hbar\, h,
$$ 
expanding the action $S(\varphi)$ in functional Taylor 
series in quantum fields $h$, expanding both sides of 
the equation in powers on $\hbar$   and equating the 
coefficients of equal powers of $\hbar$, one gets the recurrence
relations that uniquely define all coefficients $\Gamma_{(k)}$. 
All functional integrals appearing in this expansion have the form 
\bes
\int\limits_{\cal M}{\cal D}h\exp\left(-\frac{i}{2}h\cdot
\Delta h\right)h^{i_{1}}\cdots h^{i_{n}}\,,
\ees
where $\Delta$ is a partial differential operator  defined by the second
variation of the action
\bes
\Delta=-{\delta^2 S\over\delta\varphi^2}\,.
\ees
These integrals are Gaussian and can be calculated in terms 
of the functional determinant of the operator $\Delta$ and  the bare 
propagator $G=\Delta^{-1}$, i.e. the Green function of the operator  $\Delta$
with Feynman boundary conditions, and the local  classical vertex functions
$S_{,i_1\dots i_n}$. 

In particular, 
the one-loop effective action is determined by the functional 
determinant of the operator $\Delta$ 
\bes
\Gamma_{(1)}=-\frac{1}{2i}\log\Det\,\Delta\,.
\ees

Strictly speaking,  the Gaussian integrals are well defined for elliptic
differential operators  in terms of the functional determinants and their Green
functions.  Although the Gaussian integrals of quantum field theory are
determined  by hyperbolic differential operators with Feynman boundary
conditions  they can be well defined by means of the analytic  continuation
from the  Euclidean sector of the theory where the operators become elliptic. 
This is done by so-called Wick rotation---one replaces the real time
coordinate by a purely imaginary  one $x^0\to i\tau$ and singles out the
imaginary factor also from the  action $S\to iS$ and the effective action
$\Gamma\to i\Gamma$.  Then the metric of the spacetime manifold becomes
positive  definite and the classical action in all `nice' field theories
becomes a positive-definite functional. Then the fast oscillating Gaussian 
functional integrals become exponentially decreasing and can  be given a
rigorous mathematical meaning. 


Let us try to apply the formalism described above to a gauge field theory.
A characteristic feature of a gauge field theory is the fact that 
the dynamical equations 
$$
{\delta S\over \delta\varphi}=0
$$
are not independent 
--- there  are certain identities, called N\"other identities,  between them.
This means that there are some nowhere vanishing vector fields
$$
\RR_\alpha=R^i{}_\alpha{\delta\over \delta\varphi^i}
$$
on the configuration space 
${\cal M}$ that annihilate the action  
$$
{\RR}_{\alpha}S=0,
$$
and, hence,
define invariance flows on ${\cal M}$. The transformations of the fields 
$$
\delta_{\xi}\varphi^i=R^i{}_{\alpha}\xi^\alpha
$$
 are called the invariance
transformations and $\RR_{\alpha}$ are called the generators of invariance
transformations. The infinitesimal parameters of these transformations $\xi$
are sections of another vector bundle (usually the tangent bundle $TG$ of a
compact Lie group $G$) that are respresented locally by a set of functions
$$
(\xi^\alpha)=(\xi^a(x)),
$$
$a=1,\dots,\dim G$,  over spacetime with compact
support.  To distinguish between the components of the gauge fields and the
components of the gauge parameters we introduce  lower case Latin indices from
the beginning of the alphabet; the Greek indices from the beginning of the
alphabet are used as condensed labels 
$$
\alpha =(a,x)
$$
that include the
spacetime point.  We assume that the vector fields $\RR_{\alpha}$ are linearly
independent and complete, which means that they form a complete basis in the
tangent space of the invariant subspace of configuration space. The vector
fields $\RR_{\alpha}$ form the gauge algebra. We restrict ourselves to the
simplest case when the gauge algebra is the Lie algebra of an
infinite-dimensional gauge Lie group $\cal G$, which is the case in Yang-Mills
theory and gravity. Then the flow vectors $\RR_{\alpha}$ decompose the
configuration space into the invariants subspaces of ${\cal M}$ (called the
orbits) consisting of the points connected by the gauge transformations. The
space of orbits is then ${\cal M}/{\cal G}$.  The linear independence of the
vectors $\RR_{\alpha}$  at each point implies that each orbit is a copy of the
group manifold. One can show that the vector fields $\RR_{\alpha}$ are tangent
to the dynamical subspace ${\cal M}_0$, which means that the orbits do not
intersect ${\cal M}_0$ and the invariance flow maps the dynamical subspace
${\cal M}_0$ into itself. Since all  field configurations connected by a gauge
transformation, i.e., the points on an orbit, are physically  equivalent, the
physical dynamical variables  are the classes of gauge equivalent field
configurations, i.e.,  the orbits. The physical configuration space is, hence,
the  space of orbits ${\cal M}/{\cal G}$. In other words  the physical
observables must be the invariants of the  gauge group.

To quantize a gauge theory by means of the functional integral,  we consider
the in-- and out-- regions,  define some $\left.|{\rm in}\right>$ and
$\left.|{\rm out}\right>$  states in these regions and study the amplitude 
$\left<{\rm out}|{\rm in}\right>$. Since all field configurations along  an
orbit are  physically equivalent we have to integrate  over the orbit space
${\cal M}/{\cal G}$.  To deal with such situations one has to choose a
representative  field in each orbit. This can be done by choosing special
coordinates  $(I^A(\varphi),\chi^{\alpha}(\varphi))$ on the
configuration space  ${\cal M}$, where $I^A$ label the orbits and
$\chi^{\alpha}$ the points in the orbit. Computing the Jacobian of the
field transformation and introducing a delta functional $\delta(\chi-\zeta)$
we can fix the coordinates on the orbits and obtain the measure on the orbit
space ${\cal M}/{\cal G}$  
\bes
{\cal D}I={\cal D}\varphi\,\Det F(\varphi)\delta(\chi(\varphi)-\zeta),
\ees
where 
$$
F^\beta{}_\alpha=\RR_\alpha \chi^\beta
$$
is a non-degenerate operator. 
Thus we obtain a functional integral for the transition amplitude
$$
\left<{\rm out}|{\rm in}\right> 
= \int_{\cal M}{\cal D}\varphi
\Det F(\varphi)\delta(\chi(\varphi)-\zeta)
\exp\left\{{i\over\hbar}S(\varphi)\right\}.
$$
Now one can go further and integrate this equation over parameters $\zeta$ with
a Gaussian measure determined by a nondegenerate matrix
$\gamma=(\gamma_{\alpha\beta})$, which most naturally can be choosen as the
metric on the orbit (gauge group metric). As a result we get 
$$
\left<{\rm out}|{\rm in}\right>
=\int\limits_{\cal M}{\cal D}\varphi\,
\Det^{1/2}\gamma\,\Det F(\varphi)
\exp\left\{{i\over\hbar}\left[S(\varphi)
+\frac{1}{2}\chi(\varphi)\cdot
\gamma\chi(\varphi)\right]\right\}.
$$
The functional equation for the effective action takes the form
\bea
\exp\left\{{i\over \hbar}\Gamma(\phi)\right\}
&=&
  \int\limits_{\cal M} {\cal D}\varphi\,\Det^{1/2}\gamma\, \Det F(\varphi)
\nonumber\\[10pt]
&   &
\times \exp\Biggl\{{i\over \hbar}\Bigl[S(\varphi) 
+{1\over 2}\chi(\varphi)\cdot \gamma\chi(\varphi)
-{\delta\Gamma(\phi)\over\delta\phi}\cdot(\varphi-\phi)\Bigr]\Biggr\}\,.
\nonumber
\eea
This equation can be used to construct the semi-classical perturbation theory
in powers of the Planck constant (loop expansion), which gives the
effective action in terms of the bare propagators and the vertex functions.
The new feature is though that the bare propagator and the vertex functions
are determined by the action
\bea
S_{\rm eff}(\varphi) &=& S(\varphi) 
+{1\over 2}\chi(\varphi)\cdot\gamma\chi(\varphi)
\nonumber
\\[10pt]
&&
+{\hbar\over i}\log\Det F(\varphi)+{\hbar\over 2i}\log\Det\gamma
\nonumber
\eea
In particular, one finds the one-loop effective action  
\bea
\Gamma_{(1)} &=& -{1\over 2i }\log\Det\Delta
+{1\over i}\log\Det F
+{1\over 2i}\log\Det\gamma\;,
\nonumber
\eea
where  
\bes
\Delta = -{\delta^2 S\over\delta\varphi^2} 
-{\delta\chi\over\delta\varphi}\cdot \gamma\,
{\delta\chi\over\delta\varphi}\;. 
\ees


\section{Heat Kernel Asymptotic Expansion}

As we have seen in the previous lecture the effective action in quantum field
theory can be computed within the semi-classical perturbation theory---the
one-loop effective action is determined by the functional determinants of
second-order hyperbolic partial differential operators with Feynman boundary
conditions and the higher-loop approximations are determined in terms of the
Feynman propagators and the classical vertex functions. As we noted above these
expressions are purely formal and need to be regularized and renormalized,
which can be done in a consistent way in renormalizable field theories. One
should stress, of course, that many physically interesting theories (including
Einstein's general relativity) are 
perturbatively non-renor\-ma\-li\-zable. Since we only need
Feynman propagators we can do the Wick rotation and consider instead of
hyperbolic operators the elliptic ones. The Green functions of elliptic
operators and their functional determinants can be expressed in terms of
the heat kernel. That is why we concentrate in the subsequent lectures on the
calculation of the heat kernel.

The gauge invariance (or covariance) in  quantum gauge field theory and quantum
gravity is of fundamental importance. That is why, a calculational scheme
that is manifestly covariant is an inestimable advantage. A manifestly
covariant calculus is such that every step is expressed in terms of geometric
objects; it does not have some intermediate ``non-covariant'' steps that lead
to an ``invariant'' result. Below we describe a manifestly covariant method
for calculation of the heat kernel following mainly our papers
\cite{avramidi00a,avramidi91b,avramidi99,avramidi95}.

Let $(M,g)$  be a smooth compact Riemannian manifold of dimension $m$ without
boundary, equipped with a positive definite Riemannian metric
$g$. Let $V$ be a vector bundle over $M$, $V^*$ be its dual, and $\End(V)\cong
V\otimes V^*$ be the corresponding bundle of endomorphisms. Given any vector
bundle $V$, we denote by $C^\infty(V)$ its space of smooth sections.  We
assume that the vector bundle $V$ is equipped with a Hermitian metric.  This
naturally identifies the dual vector bundle $V^*$ with $V$, and defines a
natural $L^2$ inner product and the  $L^2$-trace $\Tr_{L^2}$ using the
invariant Riemannian measure on the manifold $M$.  The completion of
$C^\infty(V)$ in this norm defines the Hilbert space $L^2(V)$ of square
integrable sections. We denote by $TM$ and $T^*M$ the tangent and cotangent
bundles of $M$. Let a  connection, $\nabla^V: C^\infty(V)\to
C^\infty(T^*M\otimes V)$, on the vector bundle $V$ be given,  which we assume
to be compatible with the Hermitian metric on the vector bundle $V$. The
connection is given its unique natural extension to bundles in the tensor
algebra over $V$ and $V^*$.  In fact, using the Levi-Civita connection
$\nabla^{\rm LC}$ of the metric $g$ together with $\nabla^V$, we naturally
obtain connections on all bundles in the tensor algebra over $V,\,V^*,\,TM$
and $T^*M$; the resulting connection will usually be denoted just by $\nabla$.
It is usually clear which bundle's connection is being referred to, from the
nature of the section being acted upon. Let $\nabla^*$ be the formal adjoint
to $\nabla$ defined using the Riemannian metric and
the Hermitian structure  on $V$ and let $Q\in C^\infty(\End(V))$ be a smooth
Hermitian section of the endomorphism bundle $\End(V)$. 

A Laplace type
operator $F: C^\infty(V)\to C^\infty(V)$ is a partial differential operator
of the form  
\be 
F=\nabla^*\nabla+Q=-g^{\mu\nu}\nabla_\mu\nabla_\nu+Q\,.
\label{1ms}
\ee
It is obviously symmetric, i.e.  
$$
(F\varphi,\psi) = (\varphi,F\psi),
$$
elliptic, and can be made essentially self-adjoint, i.e. its closure is
self-adjoint, which implies that it has a unique self-adjoint
extension. We will not be very careful about this   and will simply say that
$F$ is elliptic and self-adjoint. 
It is well
known \cite{gilkey95} that: 
\begin{itemize}
\item[i)] the operator $F$ has a discrete real spectrum,
$\{\lambda_n\}_{n=1}^\infty$, bounded from below: \bes
\Lambda< \lambda_1 < \lambda_2
< \cdots<\lambda_n<\cdots
\ees
with some real constant $\Lambda$, 
\item[ii)]
all
eigenspaces of the operator $F$ are finite-dimensional, and iii) the
eigenvectors, $\{\varphi_n\}_{n=1}^\infty$, of the operator $F$, are smooth
sections of the vector bundle $V$ that form a complete orthonormal basis
in $L^2(V)$.
\end{itemize}

{}For $t>0$ the operators 
$$
U(t)=\exp(-tF)
$$
form a semi-group of bounded
operators on $L^2(V)$, so called heat semi-group. The kernel of this operator
is defined by  
\bes 
U(t|x,x')
=\sum\limits_n e^{-t\lambda_n}\varphi_n(x)\otimes \varphi^*_n(x'),
\ees
where each eigenvalue is counted with multiplicities.
It is a section of the external tensor product of vector bundles $V\boxtimes
V^*$ over $M\times M$, which 
can also be regarded as an endomorphism from the fiber of $V$ over $x'$ to the
fiber of $V$ over $x$. This kernel satisfies the heat equation 
\be
\left(\partial_t+F\right)U(t)=0
\label{he-5/01}
\ee
with the initial condition
\be
U(0^+|x,x')=\delta(x,x')\,
\label{init-5/01}
\ee
and is called the heat kernel.

Moreover, the heat semigroup $U(t)$ is a trace-class operator  with a well
defined $L^2$-trace
\be
\Tr_{L^2}\exp(-tF)=\int\limits_M \tr_V U^{\rm diag}(t)\,.
\nonumber
\ee
Hereafter
$\tr_V$ denotes the fiber trace and the label `${\rm diag}$' means the diagonal
value of a two-point quantity, e.g.  
\bes
U^{\rm diag}(t|x)=U(t|x,x')\Big|_{x=x'}\,.
\ees
The trace of the heat kernel is obviously a spectral invariant of the
operator $F$. It determines other spectral functions by integral transforms.
Of particular importance is the so-called zeta function, which enables one to
define, in particular, the regularized functional determinant of an elliptic
operator. 

In these lectures we will study the heat kernel only locally, i.e. in the
neighbourhood of the diagonal of $M\times M$, when the points $x$ and $x'$ are
close to each other. We fix a point $x'$ of the manifold and consider a small
geodesic ball with a radius smaller than the injectivity radius of the
manifold, so that each point $x$ of the ball can be connected by a unique
geodesic with the point $x'$. Such geodesic ball can be covered by a single
coordinate patch with normal local coordinates centered at $x'$.
Let $\sigma(x,x')$ be the geodetic interval, 
defined as one half the square of the length of the geodesic
connecting the points $x$ and $x'$, i.e. 
$$
\sigma(x,x')={1\over 2}[\dist(x,x')]^2.
$$
The first derivatives of this function with respect to $x$ and $x'$ define
tangent vector fields to the geodesic at the points $x$ and $x'$
\bea
u^\mu(x,x') &=& g^{\mu\nu}\partial_\nu\sigma,\quad
\nonumber\\[10pt]
u^{\mu'}(x,x') &=& g^{\mu'\nu'}\partial'_{\nu'}\sigma ,
\label{tan-5/01}
\eea
and the determinant of the mixed second derivatives defines the so-called Van
Vleck--Morette determinant 
\be
\Delta(x,x')
=|g(x)|^{-{1\over 2}}|g(x')|^{-{1\over 2}}
\det(-\partial_\mu\partial'_{\nu'}\sigma)\,.
\nonumber
\ee
Let, finally, ${\cal P}(x,x')$ denote the
parallel transport operator of sections of the vector bundle $V$ along the
geodesic from the point $x'$ to the point $x$. It  is an endomorphism from
the fiber of $V$ over $x'$ to the fiber of $V$ over $x$ (or a section of the
external tensor product $V\boxtimes V^*$ over $M\times M$). Near the diagonal
of $M\times M$ all these two-point functions are smooth single-valued
functions of the coordinates of the points $x$ and $x'$. We should point out
from the beginning that we will construct all two-point geometric
quantities (in particular, the coefficients of the asymptotic expansion of
the heat kernel as $t\to 0$) in form of covariant Taylor series. The Taylor
series do not necessarily converge in smooth case; they do, however, converge
in the analytic case in a sufficiently small neighborhood of the diagonal.

Further, one can easily prove that the function
\bes
U_0(t)  =  (4\pi t)^{-m/2}\Delta^{1/2}
\exp\left(-{\sigma\over 2t}\right)
{\cal P}
\ees
satisfies the initial condition (\ref{init-5/01}).
Moreover, locally it also satisfies the heat equation (\ref{he-5/01}) in the
free case, when the Riemannian curvature ${\rm Riem}$ of the manifold, the
curvature  ${\cal R}$ of the connection $\nabla^V$, and the endomorphism $Q$
vanish: 
$$
{\rm Riem}={\cal R}=Q=0.
$$
Therefore, $U_0(t)$ is the exact heat
kernel for the Laplacian in flat Euclidean space with a flat
trivial bundle connection.
This function gives a good framework for the approximate solution in the
general case. Namely, by factorizing out the free factor we get an ansatz
\bea
U(t) = (4\pi t)^{-{m/2}}\Delta^{{1/2}}
\exp\left(-{\sigma\over 2t}\right)
{\cal P}\,
\Omega(t)\,.
\label{150}
\eea
The function $\Omega(t|x,x')$, called the transport function, is a section of
the endomorphism bundle $\End(V)$ over the point $x'$. It  satisfies the
transport equation 
\bes
\left(\partial_t+{1\over t}D+L\right)\Omega(t)=0,
\ees
with the initial condition
\bes
\Omega(0|x,x')=\II\,,
\ees
where $\II$ is the identity endomorphism of the bundle $V$ (we will often omit
it) and $D$ and $L$ are differential operators defined by \be
D=u^\mu\nabla_\mu,
\ee
\be
L={\cal P}^{-1}\Delta^{-1/2}F\Delta^{1/2}{\cal P}\,.
\label{160}
\ee

Now, let us fix a sufficiently large negative parameter $\lambda$, viz.
$\lambda<\Lambda$, so that $(F-\lambda \II)$ is a positive operator. Since
$$
\exp[-t(F-\lambda \II)]=e^{t\lambda}\exp[-tF],
$$
the transport function for
the operator $(F-\lambda \II)$ is $e^{t\lambda}\Omega(t)$. Clearly, for
sufficiently large negative $\lambda$, $\lambda<<0$, the function
$e^{t\lambda}\Omega(t)$ with all its derivatives decreases faster than any
power of $t$ as $t\to\infty$.  Let us consider a slightly modified version of
the Mellin transform of the function $e^{t\lambda}\Omega(t)$
\be
b_q(\lambda)={1\over \Gamma(-q)}\int_0^\infty dt\,
t^{-q-1}e^{t\lambda}\Omega(t)\,. 
\label{41-7/98}
\ee
Note that for fixed $\lambda$ this is a Mellin transform of
$e^{t\lambda}\Omega(t)$ and for a fixed $q$ this is a Laplace transform of
the function $t^{-q-1}\Omega(t)$. The integral (\ref{41-7/98}) converges for
${\rm Re}\, q<0$. By integrating by parts and analytical continuation
one can prove that  the function $b_q(\lambda)$ is an entire function of $q$.
The values of the function $b_q(\lambda)$ at the integer positive points
$q=k$ are given by  
\bea
b_k(\lambda) = \sum_{n=0}^k {k\choose n}(-\lambda)^{k-n}a_{n}\,, 
\label{43-7/98}
\eea
where
\be
a_k=(-\partial_t)^k\Omega(t)\Big|_{t=0}\,.
\ee
These coefficients $a_k=a_k(x,x')$ are called 
Hadamard--Minakshisundaram--DeWitt--Seeley (HMDS) coefficients.

By inverting the Mellin transform we obtain a new ansatz for the transport
function
and, hence, for the heat kernel
\bes
\Omega(t)={1\over 2\pi i}\int\limits_{c-i\infty}^{c+i\infty}dq\,
e^{-t\lambda}t^q\,\Gamma(-q)b_q(\lambda)\,,
\ees
where $c<0$ and ${\rm Re}\,\lambda<\Lambda$. Clearly, since the left-hand
side of this equation does not depend on $\lambda$, neither does the right
hand side. Thus, $\lambda$ serves as an auxiliary parameter that regularizes
the behavior at $t\to \infty$. 

Substituting this ansatz into the transport equation we get a
functional-differential equation for the function $b_q$
\bes
\left(1+{1\over q}D\right)b_q(\lambda)=(L-\lambda I)\,b_{q-1}(\lambda)
\ees
with the initial condition
\bes
b_0(\lambda)=\II.
\ees
Note that for integer $q=k$ and $\lambda=0$ this becomes a differential
recursion system for the coefficients $a_k$
\bea
a_0 &=& \II, \qquad \\[10pt]
\left(1+{1\over k}D\right)a_k &=& L\,a_{k-1}\,.
\label{400-7/98}
\eea
It is interesting to note that there is an asymptotic expansion of
$b_q(\lambda)$ as $\lambda\to -\infty$ 
\bes
b_q(\lambda)
\sim\sum_{n=0}^\infty
{\Gamma(q+1)\over n!\Gamma(q-n+1)}(-\lambda)^{q-n}a_{n}\,,
\ees
that coincides with (\ref{43-7/98}) for integer $q$.

By computing the inverse Mellin transform we obtain the asymptotic
expansion of the transport function as $t\to 0$ in terms of the coefficients
$a_k$
\be
\Omega(t)  \sim   
\sum\limits_{k=0}^\infty{(-t)^k\over k!}a_k\,.
\label{600}
\ee

Using our ansatz (\ref{150}) we also find the trace of the heat
kernel in form of an inverse Mellin transform
\bes
\Tr_{L^2}\exp(-tF) = (4\pi t)^{-m/2}e^{-t\lambda}
{1\over 2\pi i}
\int\limits_{c-i\infty}^{c+i\infty}dq\,t^q\,\Gamma(-q)B_q(\lambda),
\ees
where
\be
B_q(\lambda)=\int\limits_M \tr_V\,b^{\rm diag}_q(\lambda)\,.
\ee
Noting that $B_q$ is an entire function of $q$, this gives the standard
asymptotic expansion as $t\to 0$ 
\bea
\Tr_{L^2}\exp(-tF) \sim
\sum\limits_{k=0}^\infty t^{(k-m)/2} A_k\,,
\label{71-5/01}
\eea
where
\be
A_{2k+1}=0
\nonumber
\ee
and
\be
A_{2k}=(4\pi)^{-m/2}{(-1)^k\over k!}\int\limits_M \tr_V\,a^{\rm diag}_k\,.
\nonumber
\ee
This is the famous Minakshisundaram--Pleijel asymptotic expansion, which is
called Schwinger--DeWitt expansion in the physics literature.
This expansion is of great importance in
differential geometry, spectral geometry, quantum field theory and other areas
of mathematical physics, such as theory of Huygens' principle, heat kernel
proofs of the index theorems, Korteveg--De Vries hierarchy, Brownian motion
etc..


The (off-diagonal) HMDS coefficients $a_k$ are determined by the recursion
system (\ref{400-7/98}).  The formal solution of
this recursion system is  
\be
a_k=D_k^{-1}\; L\; D_{k-1}^{-1}\; L \cdots D_{1}^{-1}\; L\;\II\,,  
\ee
where 
\be
D_k=1+{1\over k}D\,.
\label{Dk-5/01}
\ee

To give a precise  meaning to this formal operator solution we need  to
define the inverse operator $D_k^{-1}$. This can be done in terms of the
covariant Taylor series. We will need the following notions from the theory of
symmetric tensors.
Let $S^n_m$ be the bundle of symmetric  tensors  of type
$(m,n)$. First of all, we define the exterior symmetric tensor product
$$
\vee: \ S^n_m\times S^i_j\to S^{n+i}_{m+j}
$$
of symmetric tensors by  
\be
(A\vee B)_{\alpha_1\dots \alpha_{m+j}}^{\beta_1\dots \beta_{n+i}}
=A^{(\beta_1\dots \beta_n}_{(\alpha_1\dots \alpha_m}
B^{\beta_{n+1}\dots \beta_{n+i})}_{\alpha_{m+1}\dots \alpha_{m+j})}\,.
\label{vee-5/01}
\ee
This naturally leads to the following definition of the exterior symmetric
power of a symmetric tensor 
$$
\vee^k:\ S^n_m\to S^{nk}_{mk}
$$
\be
\vee^k A=\underbrace{A\vee\cdots\vee A}_{k}\,.
\ee
Next, we define the inner product 
$$
\star:\ S^n_m\times S^i_n\to S^i_{m}
$$
by
\be
(A\star B)_{\alpha_1\dots \alpha_{m}}^{\beta_1\dots \beta_{i}}
=A^{\gamma_1\dots \gamma_n}_{\alpha_1\dots \alpha_m}
B^{\beta_{1}\dots \beta_{i}}_{\gamma_1\dots \gamma_n}\,.
\ee
We also define the exterior symmetric covariant derivative 
$$ \nabla^S:\S^m_n\to S^m_{n+1}$$ 
by
\be
(\nabla^S A)_{\alpha_1\dots \alpha_{n+1}}^{\beta_1\dots \beta_m}
=\nabla_{(\alpha_1}A^{\beta_1\dots \beta_m}_{\alpha_2\dots \alpha_{n+1})}\,. 
\ee
These definitions are naturally extended to $\End(V)$-valued symmetric
tensors, i.e. to the sections of the bundle $S^m_n\otimes\End(V)$.


Let us consider the space of smooth two-point functions in a small
neighborhood of the diagonal $x=x'$ that we will denote by $\left.|f\right>$.
Let us define a special set of such functions $\left.|n\right>$,  labeled by
a non-negative integer $n$, by 
\bea
|0\rangle &=& 1,\\[10pt]
|n\rangle &=& {(-1)^n\over n!}\vee^n u'\,,
\eea
where $u'$ is the tangent vector field to the geodesic connecting the points
$x$ and $x'$ at the point $x'$ defined by (\ref{tan-5/01}). 
It is easy to show that these functions satisfy the equation
\be
D|n\rangle =n|n\rangle
\ee
and, hence, are the eigenfunctions of the operator $D$ with positive integer
eigenvalues. 

Let $\langle n|$ denote the dual functions defined by   
\be
\langle n|f\rangle=(\nabla^S)^nf\Big|_{x=x'}\,,
\ee
so that
\be
\langle n|m\rangle = \delta_{mn} \II_{(n)}\,,
\ee
where $\II_{(n)}$ is the identity endomorphism on the space of symmetric
$n$-tensors.
Using this notation the covariant Taylor series for an analytic function
$|f\rangle$ can be written in the form 
\be
|f\rangle=\sum_{n=0}^\infty |n\rangle\star \langle n|f\rangle\,,
\ee
and, therefore, the functions  $|n\rangle$ form a complete orthonormal basis
in the subspace of analytic functions.

The complete set of eigenfunctions $|n\rangle$ can be employed to present the
action of the operator $L$ on a function $|f\rangle$ in
the form 
\be
L|f\rangle=\sum\limits_{m,n\ge
0}|m\rangle\star\langle m|L|n\rangle \star\langle n|f\rangle, 
\label{700}
\ee
where $\langle m|L|n\rangle $ are the `matrix elements' of the operator $L$
that are just $\End(V)$-valued symmetric tensors, i.e. sections of the vector
bundle $S^n_m(M)\otimes \End(V)$.
Now it should be clear that the inverse of the operator $D_k$ in (\ref{Dk-5/01})
can be defined by
\be
D_k^{-1}|f\rangle=\sum_{n=0}^\infty{k\over k+n}|n\rangle \star\langle
n|f\rangle. 
\ee
Using such representations for the operators $D_k^{-1}$ and $L$  we obtain a
covariant Taylor series for the coefficients $a_k$ 
\be
a_k=\sum_{n=0}^\infty|n\rangle\star\langle n|a_k\rangle
\ee
where
\bea
\langle n|a_k\rangle
&=&\sum_{n_1,\dots,n_{k-1}\ge 0}
\left(\prod\limits_{j=1}^k
{j\over j+n_j}\right)
\langle n|L|n_{k-1}\rangle
\nonumber\\[13pt]
&&
\star\langle n_{k-1}|L|n_{k-2}\rangle
\star\cdots\star \langle n_1|L|0\rangle\,,
\label{800}
\eea
with $n_k\equiv n$. 

Thus, we have reduced the problem of computation of the HMDS-coefficients $a_k$ to
the evaluation of the matrix elements $\langle m|L|n\rangle$ of the
operator $L$. For a differential operator $L$ of second
order, the matrix elements $\langle m|L|n\rangle $ vanish for $n> m+2$.
Therefore, the summation over $n_i$ in (\ref{800}) is limited from above:
$n_1\ge 0$, and $n_i\le n_{i+1}+2$, for  $i=1,2,\dots,k-1$, and, hence,
the sum (\ref{800}) always contains only a finite number of terms.

We will not present here explicit formulas, (they have
been computed explicitly for arbitrary $m$, $n$ in  \cite{avramidi91b}), but
note that all these quantities are expressed polynomially in terms of three
sorts of geometric data: 
\begin{itemize}
\item[i)] symmetric tensors of type $(2,n)$, i.e. sections
of the bundle $S^2_n$ obtained by symmetric derivatives
$$
K_{(n)}=(\nabla^S)^{n-2}{\rm Riem}
$$
of the symmetrized Riemann tensor ${\rm
Riem}$ taken as a section of the bundle $S^2_2$, 
\item[ii)]
 sections  
$$
{\cal
R}_{(n)}=(\nabla^S)^{n-1}{\cal R}
$$
of the vector bundle $S^1_n\otimes \End(V)$
obtained by symmetrized derivatives of the curvature ${\cal R}$ of the
connection $\nabla^V$ taken as a section of the bundle $S^1_1\otimes\End(V)$,
\item[iii)]
$\End(V)$-valued symmetric forms, i.e. sections of the vector
bundle $S^0_n\otimes\End(V)$, constructed from the symmetrized
covariant derivatives 
$$
Q_{(n)}=(\nabla^S)^n Q
$$
 of the endomorphism $Q$. 
\end{itemize}

\section{Approximation Schemes for Calculation of the Heat Kernel}

In this lecture  we are going to investigate the general structure of the
heat kernel coefficients $A_k$. We will follow mainly our papers
\cite{avramidi91b,avramidi93b,avramidi95b,avramidi94a,avramidi96a}  (see also
our review papers \cite{avramidi95c,avramidi94b,avramidi95d}).
Our analysis will be again purely local. Since locally one can always expand
the metric, the connection and the endomorphism $Q$ in the covariant Taylor
series, they are completely characterized by their Taylor coefficients, i.e.
the covariant derivatives of the curvatures, more precisely by the objects
$K_{(n)}$, ${\cal R}_{(n)}$ and $Q_{(n)}$ defined in the previous
lecture. We introduce the following notation for all of them
$$
\Re_{(n)}=\{K_{(n+2)}, {\cal R}_{(n+1)}, Q_{(n)}\},
$$
and call these objects covariant jets; $n$ will be called the  order of a
jet $\Re_{(n)}$. It is worth noting that the jets are defined by symmetrized
covariant derivatives. This makes them well defined as ordering of the
covariant derivatives becomes not important---it is only the number of
derivatives that plays a role.

The coefficients $A_k$ are integrals of local invariants $\tr_V a_k^{\rm diag}$
which are polynomial in the jets. The first two coefficients have the well
known form
\bea
a^{\rm diag}_0 &=& \II\,,\\[10pt]
a_1^{\rm diag} &=& Q-{1\over 6}R\,,
\eea
where $R$ is the scalar curvature. For $k\ge 2$  one can classify the terms
in $A_{2k}$ according to the number of the jets and their order
\bes
A_{2k}=\sum_{n=2}^k A_{2k,(n)}\,,
\ees
where $A_{2k,(n)}$ can be presented symbolically in the form
\bea
A_{2k,(2)} &=& \int\limits_M \,{\rm tr}_V\,\sum\Re_{(0)}\Re_{(2k-4)},
\nonumber\\[10pt]
A_{2k,(3)} &=& \int\limits_M \,{\rm tr}_V\,
\sum_{i=0}^{2k-6}\sum \Re_{(0)}\Re_{(i)}\Re_{(2k-6-i)},
\nonumber\\[10pt]
A_{2k,(k-1)} &=& \int\limits_M \,{\rm tr}_V\,
\sum_{i=0}^{k-3}\sum\Re_{(0)}^i \ \Re_{(1)}\Re_{(0)}^{k-i-3}\Re_{(1)},
\nonumber\\[10pt]
A_{2k,(k)} &=& \int\limits_M \,{\rm tr}_V\,\sum\Re_{(0)}^k.
\nonumber
\eea
More precisely, all quadratic terms can be reduced to a finite number
of invariant structures, viz. \cite{avramidi91b}
\bea
A_{2k,(2)}&=&(4\pi)^{-m/2}{(-1)^k}
{(k-2)!\over 2(2k-3)!}
\nonumber\\[10pt]
&&
\times
\int\limits_M \,{\rm tr}_V\,
\Biggl\{f^{(1)}_kQ\sq^{k-2}Q
\nonumber\\[10pt]
&&
+2f^{(2)}_k{\cal R}^{\beta\gamma}\nabla_\beta
\sq^{k-3}\nabla_\alpha{\cal R}^\alpha{}_\gamma
+f^{(3)}_k Q\sq^{k-2}R
\nonumber\\[10pt]
&&
+f^{(4)}_kR^{\alpha}{}_{\beta}\sq^{k-2}R^{\beta}{}_{\alpha}
+f^{(5)}_kR\sq^{k-2}R
\Biggr\},
\nonumber
\eea
where $\sq=g^{\mu\nu}\nabla_\mu\nabla_\nu$, and $f^{(i)}_k$ are some numerical
coefficients. These numerical coefficients can be computed by the technique
developed in the previous section. The result reads
\bea
f^{(1)}_k &=& 1,\nonumber\\[10pt]
f^{(2)}_k &=& {{1}\over {2(2k-1)}},\nonumber\\[10pt]
f^{(3)}_k &=& {{k-1}\over {2(2k-1)}},\nonumber\\[10pt]
f^{(4)}_k &=& {{1}\over {2(4k^2-1)}},\nonumber\\[10pt]
f^{(5)}_k &=& {{k^2-k-1}\over {4(4k^2-1)}}.
\nonumber
\eea
One should note that the same results were obtained by a completely different
method in \cite{branson90b}.


Let us consider the situation when the curvatures are small but rapidly
varying, i.e. the derivatives of the curvatures are more important than the
powers of them. Then the leading derivative terms in the heat kernel are the
largest ones and, therefore, the trace of the heat kernel has the form
\bea
\Tr_{L^2}\exp(-tF)
& = & t^{-m/2}A_0+t^{1-m/2}A_2
\nonumber\\[10pt]
&& 
+{t^{2-m/2}}H(t) +O(\Re^3),
\nonumber
\eea
where $H(t)$ is some complicated nonlocal functional
that has the following asymptotic expansion as $t\to 0$
\bes
H(t)\sim \sum_{k=2}^\infty{t^{k-2}}A_{2k,(2)}.
\ees
Using the results for $A_{2k,(2)}$ one can easily construct such a functional
$H$ just by a formal summation of leading derivatives
\bea
H(t) &=& (4\pi)^{-m/2}{1\over 2}\int\limits_M \,\tr_V\,\Biggl\{
Q\gamma^{(1)}(-t\sq)Q
\nonumber
\\[10pt]
&&
+2{\cal R}^\alpha{}_{\gamma}\nabla_\alpha{{1}\over {\sq}}
\gamma^{(2)}(-t\sq)\nabla_\beta{\cal R}^{\beta\gamma}
-2Q\gamma^{(3)}(-t\sq)R
\nonumber
\\[10pt]
&&
+R^\alpha{}_{\beta}\gamma^{(4)}(-t\sq)R^{\beta}{}_\alpha
+R\gamma^{(5)}(-t\sq)R
\Biggr\},
\eea
where $R_{\alpha\beta}$ is the Ricci tensor, and
$\gamma^{(i)}(z)$ are some entire functions defined by
\cite{avramidi91b}
\bes
\gamma^{(i)}(z) = 
\int\limits_0^1 d\xi\, f^{(i)}(\xi)
\exp\left(-{{1-\xi^2}\over {4}}z\right)\,,
\nonumber
\ees
where
\bea
f^{(1)}(\xi) &=& 1, \nonumber\\[10pt]
f^{(2)}(\xi) &=& {{\xi^2}\over {2}}, \nonumber\\[10pt]
f^{(3)}(\xi) &=& {{1-\xi^2}\over {4}},\nonumber\\[10pt]
f^{(4)}(\xi) &=& {{1}\over {6}}\xi^4, \nonumber\\[10pt]
f^{(5)}(\xi) &=& {{1}\over {48}}(3-6\xi^2-\xi^4).\nonumber
\eea
Therefore, $H(t)$ can be regarded as generating functional for quadratic
terms $A_{2k,(2)}$ (leading derivative terms) in all coefficients $A_{2k}$.
It plays a very important role in investigating the nonlocal structure of
the effective action in quantum field theory in so-called high-energy
approximation \cite{avramidi91b}.


Let us consider now the opposite case, when the curvatures are strong but
slowly varying, i.e. the powers of the curvatures are more important than the
derivatives of them.  Since the derivatives are naturally identified with the
momentum (or energy), a situation when the derivatives are small is called the
low-energy approximation in physics literature. The investigation of the
low-energy effective action is of great importance in quantum gravity and
gauge theories because it describes the dynamics of the vacuum state of the
theory. The main terms in this approximation are the terms
without any covariant derivatives of the curvatures, i.e. the lowest order
jets. We will consider mostly the zeroth order of this approximation
which corresponds simply to covariantly constant background curvatures
\bes
\na {\rm Riem} = 0,\qquad \na {\cal R}=0,
\qquad \na Q = 0.
\ees

The trace of the heat kernel has then the form
\bes
\Tr_{L^2}\exp(-tF)=t^{-m/2}\Theta(t)+O(\nabla\Re)\,,
\ees
where $\Theta(t)$ is a functional that has the following asymptotic expansion
as $t\to 0$
\bes
\Theta(t)\sim\sum_{k=0}^\infty{t^k }A_{2k,(k)}\,.
\ees
where $A_{2k,(k)}$ are the terms without covariant derivatives (highest order
terms in the jets) in the coefficients $A_{2k}$ and $O(\nabla\Re)$ denotes the
terms with at least one derivative that vanish in the covariantly constant
case. The terms $A_{2k,k}$ are just polynomials in the curvatures and the
endomorphism $Q$. Therefore, the functional $\Theta(t)$ is a generating
functional for all heat kernel coefficients $A_k$ for a covariantly constant
background, in particular, for all symmetric spaces.


There is a very elegant indirect way to construct the heat
kernel without solving the heat equation but using only the
commutation relations of differential
operators \cite{avramidi93b,avramidi95b,avramidi94a,avramidi96a}.
The main idea is in a generalization of the usual Fourier
transform to the case of operators and consists in the following.
Let us consider for a moment a trivial case, where the
curvatures vanish but the potential term does not:
\bes
{\rm Riem}= 0,\qquad {\cal R}=0,\qquad \nabla Q=0.
\ees
In this case the operators of covariant derivatives obviously
commute and form an Abelian Lie algebra, i.e. 
$$
[\na_\mu,\na_\nu]=0.
$$
It is easy to show that the heat semigroup operator can
be presented in the form
\bea
\exp(-t F) &=& (4\pi t)^{-m/2}\exp(-tQ)
\nonumber\\[10pt]
&&\times
\int\limits_{\RR^m} dk\,|g|^{1/2}
\exp\left(-{\langle k,k\rangle\over 4t}
+k\cdot\na\right),
\nonumber
\eea
where $\langle k,k\rangle=k^\mu g_{\mu\nu}k^\nu$ and $k\cdot\nabla=k^\mu
\nabla_\mu$. Here, of course, it is assumed that the covariant derivatives also
commute with the metric, i.e. 
$$
[\nabla, g]=0.
$$
Acting with this operator on the Dirac distribution and using the
obvious relation
\be
\exp(k\cdot\na)\delta(x,x')\Big\vert_{x=x'}=\delta(k),
\label{44-5/01}
\ee
one integrates easily over $k$ and obtains the diagonal of the heat kernel
and the trace
\bes
\Tr_{L^2}\exp(-tF)=(4\pi t)^{-m/2}
\int\limits_M \tr_V\exp(-tQ).
\ees


Let us consider now a more complicated case when there is a nontrivial 
covariantly constant curvature $\R\ne 0$ in flat space: 
\bes
{\rm Riem}= 0,\qquad \nabla{\cal R}=0,
\qquad \nabla Q=0.
\ees
In this case the covariant derivatives form a nilpotent Lie algebra
\bes
[\nabla_\mu,\nabla_\nu]={\cal R}_{\mu\nu}\,,
\ees
\bes
[\nabla_\mu,{\cal R}_{\alpha\beta}]=[\nabla_\mu,Q]=0\,,
\ees
\bes
[{\cal R}_{\mu\nu},{\cal R}_{\alpha\beta}]=[{\cal R}_{\mu\nu},Q]=0\,.
\ees
For this algebra one can prove a theorem expressing the heat
semigroup operator in terms of an average over the
corresponding Lie group
\cite{avramidi93b}
\bea
\exp(-t F) &=& (4\pi t)^{-m/2}\exp(-tQ)
\det^{1/2}_{\End(TM)}\left({t\h R\over \sinh(t\h R)}\right)
\nonumber\\[10pt]
&&\times\int\limits_{\RR^m}dk\,|g|^{1/2}
\exp\left[-{1\over 4t}\langle k, t\h R \coth(t\h R) k\rangle\right]
\exp\left(k\cdot\nabla\right)\,.
\nonumber
\eea
Here functions of the curvature $\h R$ are understood as
functions of sections of the bundle $\End(TM)\otimes\End(V)$,
and the determinant
$\det_{\End(TM)}$ is taken with respect to $\End(TM)$ indices,
$\End(V)$ indices being intact.

It is not difficult to show that in this case the equation (\ref{44-5/01}) is
still valid, so that the integral over $k^\mu$ becomes trivial and
we obtain immediately the trace of the heat kernel \cite{avramidi93b}
\bea
\Tr_{L^2}\exp(-tF) &=& (4\pi t)^{-m/2}
\int\limits_M \tr_V\exp(-tQ)
\det^{1/2}_{\End(TM)}\left({t\h R \over \sinh(t\h R)}\right). 
\nonumber
\eea
Expanding this in a power series in $t$ one can find all
covariantly constant terms $A_{k,k}$ in all heat kernel coefficients $A_k$.

Let us now generalize the algebraic approach to the case of curved
manifolds with covariantly constant Riemann curvature and trivial 
connection $\nabla^V$  
\bes
\nabla {\rm Riem}=0, \qquad \R=0, \qquad \nabla Q=0.
\ees

First of all, we give some definitions. The condition $\nabla {\rm Riem}=0$
defines the geometry of locally symmetric spaces. A Riemannian locally
symmetric space which is simply connected and complete is a globally
symmetric space (or, simply,  symmetric space). A symmetric space is said to
be of  compact, noncompact or Euclidean type if all sectional curvatures 
$K(u,v)=R_{\alpha\beta\gamma\delta}u^\alpha v^\beta u^\gamma v^\delta$ are
positive, negative or zero. A direct product of symmetric spaces of  compact
and noncompact types is called semisimple symmetric space.  A generic complete
simply connected Riemannian symmetric space is a direct product of a flat
space and a semisimple symmetric space.

It should be noted that our analysis is purely local.  We are looking for a
universal (in the category of locally symmetric spaces) local  generating
function of the curvature invariants, that reproduces adequately  the
asymptotic expansion of the trace of the heat kernel. This function should
give all the terms without covariant derivatives  of the curvature $A_{2k,(k)}$ in
the asymptotic expansion of the heat kernel, i.e. in other words all heat
kernel coefficients $A_{2k}$  for any locally symmetric space. It turns
out to be much easier to obtain a universal generating function of $t$
whose Taylor coefficients reproduce the heat kernel coefficients $A_{2k}$ than
to compute them directly.

It is obvious that no flat subspace contributes to the
coefficients $A_{2k}$. Therefore, to find this universal structure it is
sufficient to consider only semisimple symmetric spaces. Moreover, since
the coefficients $A_k$ are polynomial in the curvatures, one can restrict
oneself only to symmetric spaces of compact type.  Using the factorization
property of the heat kernel and the duality  between the compact and
the noncompact symmetric spaces one can obtain then the  results for the
general case by analytical continuation. That is why we consider only the case
of  compact symmetric spaces  when the sectional curvatures and the metric are 
positive definite.

First of all, we choose a basis for the tangent bundle $TM$ that is parallel
(covariantly constant) along the geodesics. The frame components of the
curvature tensor of a symmetric space are, obviously, constant and can be
presented in the form 
\bes
R_{abcd} = \b_{ik}E^i{}_{ab}E^k{}_{cd},
\ees
where $E^i{}_{ab}$, $i=1,\dots, p$,  is some set of antisymmetric
$m\times m$ matrices, with $p$ being a constant satisfying $p\le m(m-1)/2$, and
$\b_{ik}$ is some symmetric $p\times p$ nondegenerate matrix. The traceless
matrices  $D_i=\left(D^a{}_{ib}\right)$ defined by   
\bes
D^a{}_{ib}=-\b_{ik}E^k{}_{cb}g^{ca}= - D^a{}_{bi}
\ees
are known to be the generators of the holonomy algebra ${\cal H}$
\bes
[D_i, D_k] = F^j{}_{ik} D_j,
\ees
where $F^j_{\ ik}$ are the structure constants.

In symmetric spaces a much richer algebraic structure exists. Indeed, let us
define the quantities $C^A{}_{BC}=-C^A{}_{CB}$, $A=1,\dots, D$, where $D=m+p$,
by 
\bes
C^i{}_{ab}=E^i{}_{ab}, \quad C^a{}_{ib}
=D^a{}_{ib}, \quad C^i{}_{kl}=F^i{}_{kl},
\ees
\bes
C^a{}_{bc}=C^i{}_{ka}=C^a{}_{ik}=0,
\ees
and the matrices $C_A=\left(C^B{}_{AC}\right)=(C_a,C_i)$:
\bes
C_a = \left( \matrix{ 0          & D^b{}_{ai}   \cr
		      E^j{}_{ac} & 0            \cr}\right),
\ees
\bes
C_i = \left( \matrix{ D^b{}_{ia} & 0            \cr
		      0          & F^j{}_{ik}   \cr}\right).
\ees
One can show that they satisfy the Jacobi identities
\cite{avramidi94a,avramidi96a} 
\bes
[C_A, C_B]=C^C{}_{AB}C_C
\ees
and, hence, define a Lie algebra ${\cal G}$ of dimension $D$ with the
structure constants $C^A{}_{BC}$, the matrices $C_A$ being generators of
the adjoint representation.

In symmetric spaces one can find explicitly the generators of the
infinitesimal isometries, i.e. the Killing vector fields $\xi_A$,  and show
that they form a Lie algebra of isometries that is (in case of semisimple
symmetric space) isomorphic to the Lie algebra ${\cal G}$, viz.
\bes
[\xi_A,\xi_B]=C^C{}_{AB}\xi_C\,.
\ees
Moreover, introducing a symmetric nondegenerate $D\times D$ matrix
\bes
\g_{AB} = \left(\matrix{ g_{ab} & 0             \cr
	       0                & \b_{ik}        \cr}\right),
\ees
that plays the role of the metric on the algebra ${\cal G}$, one can express
the operator $F$ in semisimple symmetric spaces in terms of the generators of
isometries 
\bes
F=-\g^{AB}\xi_A\xi_B+Q,
\ees
where $\g^{AB}=(\g_{AB})^{-1}$.

Using this representation one can prove that the heat  semigroup operator can
be presented in terms of an average over the group of isometries $G$
\cite{avramidi94a,avramidi96a} 
\bea
\exp(-t F) &=&
(4\pi t)^{-D/2}\exp\left[-t\left(Q-{1\over 6} R_G\right)\right]
\nonumber\\[10pt]
&&
\times\int\limits_{\RR^D} d k\,|\g|^{1/2}
\det^{1/2}_{{\rm Ad}({\cal G})}\left({\sinh(k\cdot C/2)\over k\cdot C/2}\right)
\nonumber\\[10pt]
&&\times
\exp\left[-{1\over 4t}\langle k,k\rangle
+k\cdot \xi\right]
\nonumber
\eea
where $|\g|=\det\g_{AB}$, \ $\langle k,k\rangle=k^A\gamma_{AB}k^B$, \ $k\cdot
C=k^A C_A$, \ $k\cdot \xi=k^A\xi_A$, and $R_G$ is the scalar curvature of the
group of isometries $G$ 
\bes
R_G= -{1\over 4}\g^{AB} C^C_{\ AD}C^D_{\ BC}.
\ees

Acting with this operator on the Dirac distribution $\d(x,x')$ one can, in
principle, evaluate the off-diagonal heat kernel $\exp(-tF)\d(x,x')$, i.e. for
non-coinciding points $x\ne x'$ (see \cite{avramidi96a}). To calculate the trace
of the heat kernel, it is sufficient to compute only the coincidence limit
$x=x'$. Splitting the integration variables $k^A=(q^a,\om^i)$ and  solving
the equations of characteristics one can obtain the action of the isometries
on the Dirac distribution  \cite{avramidi94a,avramidi96a} 
\bea
\exp\left(k\cdot\xi\right)\d(x,x')\Big\vert_{x=x'}
=\det^{-1}_{\End(TM)}\left({\sinh(\om\cdot D/2)\over
\om\cdot D/2}\right)\d(q).
\nonumber
\eea
where $\om\cdot D=\om^i D_i$.

Using this result one can easily integrate over $q$ to get the
heat kernel diagonal. After changing the integration variables $\om \to \sqrt
t\, \om$ it takes the form \cite{avramidi94a,avramidi96a} 
\bea
U^{\rm diag}(t) &=& (4\pi t)^{-m/2}\exp\left[-t\left(Q-{1\over 8}R
-{1\over 6} R_H\right)\right]
\nonumber\\[10pt]
&&\times
(4\pi)^{-p/2}\int\limits_{\RR^p}d \om\, \b^{1/2}
\exp\left(-{1\over 4} \langle\om,\om\rangle \right)
\nonumber\\[10pt]
&&\times
\det^{1/2}_{{\rm Ad}({\cal H})}\left({\sinh(\sqrt t \om\cdot F/2)
\over \sqrt t \om\cdot F/2}\right)
\nonumber\\[10pt]
&&\times
\det^{-1/2}_{\End(TM)}\left({\sinh(\sqrt t \om\cdot D/2)
\over \sqrt t \om\cdot D/2}\right),
\label{2000}
\eea
where $\beta=\det\beta_{ij}$, \ $\langle \om,\om\rangle =
\om^i\beta_{ij}\om^j$, \ $\om\cdot F = \om^i F_i$, \ $F_i=(F^j{}_{ik})$ are
the generators of the holonomy algebra ${\cal H}$ in adjoint representation
and 
\bes
R_H=-{1\over 4}\b^{ik}F^m{}_{il}F^l{}_{km}
\ees
is the scalar curvature of the holonomy group.

The remaining integration over $\om$ in (\ref{2000}) can be done in a rather
formal way \cite{avramidi94b,avramidi95d}. Let $a^{*}_i$ and $a_{k}$ be some 
operators acting on a Hilbert space that form the following $p$-dimensional
Lie algebra   
\bes 
[a^j, a^{*}_k]=\delta^j_k,
\ees
\bes
[a^i,a^k]=[a^{*}_i,a^{*}_k]=0.
\ees
Let $|0\rangle$ be the `vacuum vector' in the Hilbert space, i.e.
\bes
\langle 0|0\rangle=1,
\ees
\bes
a^i|0\rangle=0,
\ees
\bes
\langle 0|a^{*}_k=0.
\ees
Then the heat kernel (\ref{2000}) can be presented in an algebraic form
without any integration, i.e.
\bea
U^{\rm diag}(t) &=& (4\pi t)^{-m/2}
\exp\left[-t\left(Q-{1\over 8}R
-{1\over 6}R_H\right)\right]
\nonumber\\[10pt]
&&
\times\Big<0\Big|
\det^{1/2}_{{\rm Ad}({\cal H})}\left({\sinh(\sqrt t a\cdot F/2)\over
\sqrt t a\cdot F/2}\right)
\nonumber\\[10pt]
&&\times
\det^{-1/2}_{\End(TM)}\left({\sinh(\sqrt t a\cdot D/2)\over
\sqrt t a\cdot D/2}\right)
\nonumber\\[10pt]
&&
\times
\exp\left(\langle a^{*},\beta^{-1}a^{*} \rangle \right)\Big|0\Big>.
\nonumber
\eea
where $a\cdot F=a^k F_k$ and $a\cdot D=a^k D_k$. This formal solution should
be understood as a power series in the operators $a^k$ and $a^{*}_k$; it
determines a well defined asymptotic expansion as $t\to 0$.

By expanding these formulas in an asymptotic power series as $t\to 0$ one
obtains all HMDS-coefficients $a^{\rm diag}_k$ for any locally symmetric
space. Thereby one finds all covariantly constant terms $A_{2k,(k)}$ in all
heat kernel coefficients.



\section{Heat-kernel Asymptotics for Non-Laplace Type Operators}

In this lecture, we study a  general class of second-order non-Laplace
type elliptic partial differential operators, acting on sections of a
vector bundle $V$ over a Riemannian manifold $M$ without boundary following
our papers \cite{avramidi01a,avramidi01b}.  In general, the study of 
spin-tensor quantum gauge fields in a general gauge necessarily leads to
non-Laplace type operators acting on sections of  general spin-tensor bundles
described in the first lecture. It is precisely these operators that are of
prime interest in the present lecture. The study of non-Laplace operators is
quite new, and the available methods are still underdeveloped in comparison
with the Laplace type theory. The only exception to this is the case of
anti-symmetric forms, which is pretty simple and, therefore, is well understood
now \cite{gilkey91,branson94,gusynin95,gusynin99,branson97}.

So, we will restrict our attention to operators acting on tensor-spinor
bundles.  These bundles may be characterized as those appearing as  direct
summands of iterated  tensor products of the cotangent and spinor bundles, i.e.
$V=TM\otimes\cdots\otimes TM\otimes T^*M\otimes\cdots T^*M\otimes  {\cal S}$,
with ${\cal S}$ being the spinor bundle. Alternatively, they may be described
abstractly as bundles associated to representations of the spin group ${\rm
Spin}(m)$. These are extremely interesting and important bundles, as they
describe the fields in Euclidean quantum field theory.  The connection on the
tensor-spinor  bundles is built in a canonical way from the Levi-Civita
connection. The generators are determined by the representation of ${\rm
Spin}(m)$ which induces the bundle $V$; they are tensor-spinors  constructed
purely from Kronecker symbols, together with the fundamental tensor-spinor
if spin structure is involved.  More general bundles appearing in
field theory are actually tensor products of these with auxiliary bundles,
usually carrying another (gauge) group structure. 

Let $Q$ be a smooth Hermitian section of the bundle $\End(V)$, i.e. $Q^*=Q$,
and $a$ be a parallel symmetric Hermitian $\End(V)$-valued tensor, more
precisely, a smooth section of the vector bundle $TM\otimes TM\otimes\End(V)$
satisfying the following conditions 
\bea
a^{\mu\nu} &=& a^{\nu\mu},\qquad
\nonumber\\[10pt]
\left(a^{\mu\nu}\right)^* &=& a^{\mu\nu},\qquad \nonumber\\[10pt]
\N a &=& 0\,.
\label{gencond}
\eea
The operator of our primary interest in this lecture has the form
\bea
F &=& \nabla^* a\nabla+Q \nonumber\\[10pt]
  &=& -a^{\mu\nu}\nabla_\mu\nabla_\nu+Q\,.
\label{genlform}
\eea
Non-Laplace type operators appear naturally in the context of Stein-Weiss
operators \cite{branson97}. Let 
$$
T^*M\otimes V=W_1\oplus \cdots \oplus W_n
$$
 be
the decomposition of the bundle $T^*M\otimes V$ in irreducible components
$W_j$ and 
$$
{\rm Pr}_i:\ T^*M\otimes V\to W_i
$$
 be the corresponding projections.
Stein-Weiss operators 
$
G_i:\; C^\infty(V)\to C^\infty(W_i)
$
 are first
order partial differential operators (called the gradients) defined by
$$
G_i={\rm Pr}_i\nabla.
$$
Then the operator   $F:\; C^\infty(V)\to C^\infty(V)$
defined by 
\bes
F=\sum_{i=1}^n c_i G^*_iG_i\,,
\ees
with some constants $c_i$, is a second-order non-Laplace type operator of the
form (\ref{genlform}) with 
\bes
a=\sum_{i=1}^n c_i ({\rm Pr}_i)^*{\rm Pr}_i\,.
\ees
Now it is obvious that the structure of the coefficient $a$ depends solely on
the representation of the spin group with which the bundle $V$ is associated.

We will restrict ourselves to a special class when the coefficient $a$ is
built in a universal, polynomial way,  using tensor product and contraction
from the metric $g$ and its inverse $g^*$, together with (if applicable) the
volume form $\epsilon$ and/or the fundamental tensor-spinor $\g$. Such a
tensor-endomorphism $a$ is obviously parallel.  
Here we do not assume that $a^{\mu\nu}$ has the form $g^{\mu\nu}\II_V$ or
$g^{\mu\nu}B$ with some automorphism $B\in \Aut(V)$.
We do not set any conditions on the endomorphism
$Q$, except that it should be Hermitian.  


In the following we will denote the leading symbol of the operator $F$,
$\sigma_L(F;x,\xi)$, with $\xi$ a cotangent vector, just by
$A(x,\xi)$. For the non-Laplace type operator $F$ in (\ref{genlform}) it has the
form 
\bes 
A(x,\xi)=a^{\mu\nu}(x)\xi_\mu\xi_\nu\,.
\ees
We require that the leading symbol should be positive definite, i.e.
$A(x,\xi)$ is a Hermitian and positive definite endomorphism for any $(x,\xi)$,
with  $\xi\ne 0$. In particular, $F$ is elliptic. Positive definiteness
implies that the roots of the characteristic polynomial 
\bes
\chi(x,\x,\lambda):=\det_V \left[A(x,\x)-\l\right]
\ees
are positive functions on $M$.  

A very important point is that the structure of the spectrum, i.e. the number
$s$ of eigenvalues, $\lambda_1$, $\dots$, $\lambda_s$, and their multiplicities
$d_1\,,\ldots,d_s$ are constant on $M$.
Moreover, one can show that $\tr_V A^n(x,\xi)$, and, therefore, the
characteristic polynomial $\chi(x,\xi,\lambda)$ depends on $(x,\xi)$ only
trough $|\xi|^2=g^{\mu\nu}(x)\xi_\mu\xi_\nu$. As a result, the dependence of
the eigenvalues $\l_i(x,\xi)$ on $(x,\x)$ is only through $|\xi|^2$ as well. 
Since $A(x,\x)$ is $2$-homogeneous in  $\x$, the $\l_i$ must be also:
\bes
\l_i(x,\x)=\mu_i|\x|^2\,,
\ees
for some positive real numbers $\m_1\,,\ldots,\m_s$, which are independent
of the point $(x,\x)\in T^*M$, and, in fact, independent of the specific
Riemannian manifold $(M,g)$. 

Let $\Pi_i$ be the orthogonal projection onto the $\l_i\,$-eigenspace.
The $\P_i$ satisfy the conditions
\bea
\Pi_i^2 &=& \Pi_i\,,\qquad \nonumber\\[10pt]
\Pi_i \Pi_k &=& 0\,, \qquad (i\ne k)\,, \nonumber\\[10pt]
\sum_{i=1}^s \Pi_i &=& \II_{V}\,,\qquad \nonumber\\[10pt]
\tr_{V} \Pi_i &=& d_i\,. \nonumber
\eea
In contrast to the eigenvalues, the projections  depend on the  direction
$\xi/|\xi|$ of $\xi$, rather than on the magnitude $|\x|$. In other words,
they are $0$-homogeneous in $\xi$.  Furthermore, they are even polynomials in
$\xi/|\xi|$: 
\bes
\Pi_i(x,\xi)=\sum_{n=0}^{p} 
{1\over |\xi|^{2n}}\xi_{\mu_1}\cdots\xi_{\mu_{2n}}
\Pi_{i(2n)}^{\mu_1\cdots\mu_{2n}}(x)\,,
\ees
where $p$ is some positive integer. Here the $\Pi_{i(2n)}$ are some
$\End(V)$-valued trace-free symmetric $2n$-tensors that do not depend on $\xi$.
  Clearly, the leading symbol can be written in terms of eigenvalues
and projections 
\be
A(x,\x)=|\xi|^2\sum_{i=1}^s\m_i\P_i(x,\x)\,.
\label{AtoPi}
\ee
There is also a converse formula for the projections in terms of powers of
the leading symbol. Note that the highest degree of projections, $p$, is
also a constant that depends only on the representation to which $V$ is
associated; both $s$ and $p$ can be computed explicitly in
representation-theoretic terms \cite{avramidi01b}.


The non-Laplace type operator $F$ with a positive definite leading symbol is
an elliptic self-adjoint operator of second order. Therefore, there is a well
defined heat kernel $U(t|x,x')$.  Moreover, there is a well defined heat
kernel diagonal $U^{\rm diag}(t)$ and the trace of the heat kernel
$\Tr_{L^2}\exp(-tF)$ that have the asymptotic expansion of the standard form
(\ref{71-5/01}). Since the global heat kernel coefficients $A_k$ are
determined by the integrals of the fiber trace of the local ones $a^{\rm
diag}_k$, it is sufficient to compute the local heat kernel coefficients, more
precisely, their fiber traces, $\tr_V a^{\rm diag}_k$. By invariance theory,
these coefficients are linear combinations of the local invariants built from
the geometric objects (curvatures, the potential $Q$, and their derivatives)
with universal numerical constants. It is these universal constants that we
want to compute. Therefore, this can be done at any fixed point of the
manifold. 

Let us stress here that our purpose is not to provide a rigorous construction
of the heat kernel with estimates; for this we rely on the standard
references \cite{gilkey95}.  Rather, given that the existence of heat kernel
asymptotic expansion is known, our aim is to compute its coefficients. 

Our analysis will be again purely local. We fix a point $x'$ in the
manifold $M$ and consider a small geodesic ball with the radius smaller than
the injectivity radius of the manifold. Then any point in this ball can be
connected with the fixed point $x'$ by a unique geodesic. 
Further, we  represent the heat
kernel in the form 
\bea
U(t|x,y)&=&\Delta^{1/2}(x,x'){\cal P}(x,x'){\cal U}(t|x,y;x')
\nonumber\\[10pt]
&&\times
{\cal P}^{-1}(y,x')
\Delta^{1/2}(y,x')\,,
\eea
where $\Delta$ is the Van Vleck-Morette determinant and ${\cal P}$ is the
parallel transport operator defined in lecture $2$. Then the modified
heat kernel ${\cal U}$ is a section of the bundle $\End(V)$ at $x'$ but is
scalar at $x$ and $y$. It satisfies the modified heat equation \be
(\partial_t+L){\cal U}(t)=0
\ee
where $L={\cal P}^{-1}\Delta^{-1/2}F\Delta^{1/2}{\cal P}$ is the operator
defined by (\ref{160}), with the initial condition 
\bes
{\cal U}(0^+|x,y;x')=\Delta^{-1}(x,x')\delta(x,y)\,.
\ees
Here and everywhere below, as usual, the differential operators act on the
first space argument of the heat kernel (recall that $x'$ is being fixed). 

We shall employ the standard scaling device for the heat kernel ${\cal
U}(t|x,y;x')$ when $x\to x'$, $y\to x'$, and $t\to 0$. We introduce a small
expansion parameter  $\varepsilon$, choose the normal  coordinates at $x'$,
and scale the coordinates according to 
\bea
x &\to& x'+\varepsilon(x-x')\,, \qquad \nonumber\\[10pt] 
y &\to& x'+\varepsilon(y-x')\,, \nonumber\\[10pt] 
t &\to& \varepsilon^2t,
\label{scal-5/01}
\eea
Note that this also means that the derivatives scale according to 
\bea
\partial_t &\to& {1\over\varepsilon^2}\partial_t, \qquad 
\nonumber\\[10pt]
\partial_\mu &\to& {1\over\varepsilon}\partial_\mu\,.
\nonumber
\eea
%
Note also that in normal coordinates 
$$
\Delta(x,x')=|g(x)|^{-1/2},
$$
 so that
$$
\Delta^{-1}(x,x')\delta(x,y)=\delta(x-y),
$$
and 
$$
{\cal P}(x,x')=\II
$$
(however, $\nabla {\cal P}\ne 0$!).

Next, we expand the operator $L$ and the heat kernel ${\cal U}$ in a
formal asymptotic series in $\varepsilon$
\bes
L\sim \sum_{n=0}^\infty \varepsilon^{n-2}F_n\,,
\ees
and
\bes
{\cal U}(t)\sim \sum_{n=0}^\infty \varepsilon^{n-m}U_n(t)\,.
\ees
The zeroth order heat kernel is determined by the equation
\be
(\partial_t+F_0)U_0(t)=0\,
\ee
with the initial condition
\be
U_0(0^+|x,y;x')=\delta(x-y)\,.
\ee
The higher order approximations are determined by the following
differential recurrence relations
\bes
(\partial_t+F_0)U_k(t)=-\sum_{n=0}^{k-1} F_{k-n} U_{n}(t)\,
\ees
with the initial conditions
\bes
U_k(0^+|x,y;x')=0\,.
\ees
By construction the coefficients $U_n$ are homogeneous functions, i.e.
\bea
U_n(t|x,y;x') &=& t^{(n-m)/2}
\times
U_n\left(1\Big|x'+{x-x'\over\sqrt t},x'+{y-x'\over\sqrt t};x'\right)\,.
\eea
Therefore, on the diagonal one obtains the asymptotic expansion
\be
U^{\rm diag}(t)\sim\sum_{n=0}^\infty t^{(n-m)/2}U_n^{\rm diag}(1)\,,
\ee
where
\bes
U_n^{\rm diag}(t|x)=U_n(t|x,x;x)\,.
\ees
Comparing this with the standard heat kernel asymptotic expansion
(\ref{71-5/01}) we see that the diagonal odd-order coefficients vanish
\bes
U^{\rm diag}_{2k+1}(t)=0\,,
\ees
and the even-order ones give the heat kernel coefficients
\bes
A_{2k}=\int\limits_M \tr_V U^{\rm diag}_{2k}(1)\,. 
\ees

Using the form of the operator $L$ in the leading order
\be
F_0=-a^{\mu\nu}(x')\partial_\mu\partial_\nu\,
\ee
we easily find the leading order heat kernel by Fourier transform 
\bes 
U_0(t|x,y;x')=\int\limits_{\RR^m}
{d\xi\over (2\pi)^m} e^{i\xi\cdot(x-y)}\exp[-tA(x',\xi)], 
\ees 
where $\xi\cdot (x-y)=\xi_\mu (x^\mu-y^\mu)$. Here and everywhere below all
integrals over  $\xi$ will be over $\RR^m$.
Writing the leading symbol in terms of the projections,
we get
\bea
U_0(t|x,y;x') &=& \sum_{i=1}^s
\int {d\xi\over (2\pi)^m} e^{i\xi\cdot(x-y)-t\mu_i|\xi|^2}
\Pi_i(x',\xi)\,.
\nonumber
\eea
The trace of the diagonal can now be easily computed
\be
\tr_{V} U^{\rm diag}_0(t)=
\sum_{i=1}^s d_i (4\pi t\mu_i)^{-m/2}\,.
\ee
It gives the diagonal value of the lowest order heat kernel coefficient
$a_0\,$: \bes
\tr_V  a^{\rm diag}_0=\sum_{i=1}^s {d_i\over \mu_i^{m/2}}\,,
\ees
and, therefore,
\bes
A_0 = (4\pi)^{-m/2}\sum_{i=1}^s {d_i\over \mu_i^{m/2}}\vol(M)\,.
\ees
These formula points out a new  feature of non-Laplace type operators; one
which complicates life somewhat. Whereas the dimension dependence 
of the heat coefficients of Laplace type operators is isolated in the
overall factor $(4\pi)^{-m/2}$, the dimension dependence for non-Laplace
type operators is more complicated.


The calculation of higher-order coefficients is a challenging task. 
We will indicate how the coefficient $A_1$ is computed. By the invariance
theory we have,  
\bes
A_2=\int\limits_M \tr_V\left(HQ+\II\,\beta R\right)\;,
\ees
where $\beta$ is a universal constant and $H$ is some endomorphism. Both
$\beta$ and $H$ depend only on the leading symbol of the operator $F$.

To compute these quantities we need, first of all, the Taylor
expansion of the metric and connection in normal coordinates
\bea
g_{\mu\nu}(x)&=&\delta_{\mu\nu}
-{1\over 3}R_{\mu\alpha\nu\beta}(x^\alpha-x'^\alpha)(x^\beta-x'^\beta)
\nonumber\\[10pt]
&&+O[(x-x')^3],
\nonumber\\[10pt]
{\cal A}_{\mu}(x) &=& -{1\over 2}{\cal R}_{\mu\alpha}(x^\alpha-x'^\alpha)
+O[(x-x')^2]\,,
\nonumber\\[10pt]
Q(x) &=& Q+O[(x-x')].
\eea
Here and below all coefficients are computed at the fixed point $x'$; we do
not indicate this explicitly. Similarly,  the Taylor expansion of the
tensor-endomorphism $a^{\mu\nu}(x)$ is determined by the equation 
$\nabla_\mu a^{\alpha\beta}=0$,
which gives
\bea
a^{\mu\nu}(x) &=& a^{\mu\nu}
+{1\over 3}a^{\lambda(\mu}R^{\nu)}{}_{\alpha\lambda\beta}
(x^\alpha-x'^\alpha)(x^\beta-x'^\beta)
\nonumber\\[10pt]
&&
+O[(x-x')^3].
\nonumber
\eea
Using these formulas we obtain 
\bea
F_1 &=&0, \nonumber\\[10pt]
F_2&=&
X^{\mu\nu}{}_{\alpha\beta}(x^\alpha-x'^\alpha)(x^\beta-x'^\beta)
\partial_\mu\partial_\nu
\nonumber\\[10pt]
&&
+Y^\mu{}_\alpha(x^\alpha-x'^\alpha)\partial_\mu+Q,
\nonumber
\eea
where
\bea
X^{\mu\nu}{}_{\alpha\beta}
&=&
-{1\over 3}
a^{\lambda(\mu}
R^{\nu)}{}_{(\alpha|\lambda|\beta)},
\nonumber\\[10pt]
Y^\mu{}_\alpha
&=&
{2\over 3}
a^{\mu\lambda}
R_{\lambda\alpha} 
-{1\over 2}
[{\cal R}_{\alpha\nu}, a^{\mu\nu}]_+, 
\nonumber
\eea
and $[A , B]_+=AB+BA$ denotes the anticommutator.

{}From the recurrence relations (and the initial conditions) we
find 
\bes
U_1=0\,
\ees
and
\bes
U_2(t)=-\int\limits_0^t d\tau U_0(t-\tau) F_2 U_0(\tau)\,.
\ees
By using the Fourier representation for $U_0$ we obtain for the diagonal
\bea
U^{\rm diag}_2(t)
&=&
-\int{d\xi\,\over (2\pi)^{m}}
\int\limits_0^t d\tau e^{-(t-\tau)A(x,\xi)}
\hat F_2e^{-\tau A(x,\xi)}\,,
\nonumber
\eea
where
\bes
\hat F_2 =
X^{\mu\nu}{}_{\alpha\beta}\partial_\xi^\alpha 
\partial_\xi^\beta\xi_\mu\xi_\nu
-Y^\mu{}_\alpha \partial_\xi^\alpha\xi_\mu
+Q\,.
\ees
This gives finally the heat kernel coefficient $A_2$
\bea
A_2=-(4\pi)^{-m/2} \int\limits_M \int {d\xi\over \pi^{m/2}}\,\int\limits_0^1 d\tau 
\tr_V e^{-(1-\tau)A(x,\xi)}\hat F_2e^{-\tau A(x,\xi)}.
\nonumber
\eea
It is not very difficult to obtain from here the endomorphism $H$; it is given
by 
\bea
H(x) & = & -(4\pi)^{-m/2}\sum_{i=1}^s \mu^{-m/2}_i
\int {d\xi\over \pi^{m/2}} e^{-|\xi|^2}\Pi_i(x,\xi)\,.
\nonumber
\eea
The calculation of the coefficient $\beta$ is a much more complicated problem.
After a long calculation one obtains a complicated expression in terms of
the constant $\mu_i$ (the eigenvalues of the leading symbol) and the
leading symbol of the operator $F$ (see, \cite{avramidi01a}).

An interesting feature of the non-Laplace type operators is the 
semi-classical polarization. This means that, unlike Laplace type
operators, the asymptotic expansion of the heat kernel for non-Laplace type
operators has the form
\bea
U(t|x,x') &=&\sum_{i=1}^s (4\pi t\mu_i)^{-m/2}\Delta^{1/2}(x,x')
\exp\left[-{\sigma(x,x')\over 2t\mu_i}\right]\Omega_i(t|x,x')\,
\nonumber
\eea
where each transport function $\Omega_i$ satisfies a different transport
equation. This also implies that the differential recursion system
for the coefficients of the asymptotic expansion of the transport functions
is much more complicated.



\section{Heat-Kernel Asymptotics of Oblique \\
Boundary-Value
Problem}

In this lecture we study the heat kernel asymptotics for a Laplace type
partial differential operator acting on sections of a vector bundle over a
compact Riemannian manifold with boundary. In this case  one has to impose
some boundary conditions in order to make a (formally self-adjoint)
differential operator self-adjoint (at least symmetric) and elliptic. There are
many admissible boundary conditions that guarantee the self-adjointness and
ellipticity of the problem. The simplest boundary conditions are the classical
Dirichlet and the Neumann ones. There exist also slight modifications of the
Neumann boundary conditions (called Robin boundary conditions in physical
literature) when the normal derivative of the field at the boundary is
proportional to the value of the field at the  boundary
\cite{branson90a,branson99}. In an even more general scheme, called mixed
boundary conditions, the Dirichlet and Robin boundary conditions are mixed by
using some projectors. In this lecture we study a more general setup, called
the oblique boundary-value problem, which  includes  both normal and 
tangential derivatives of the fields at the boundary \cite{grubb74,gilkey83b}.
We will follow mainly our papers \cite{avramidi99a,avramidi99b}.


Let $(M,g)$ be a smooth compact Riemannian manifold of dimension $m$ with
smooth boundary $\partial M$ with a positive-definite
Riemannian metric $g$ on $M$ and induced metric $\hat g$ on $\partial
M$. In this lecture the   Greek indices range from 1 through $m$ and lower
case Latin indices range from $2$ through $m$. Let 
$$
\{\hat
e_i\}=\{\hat e_2,\dots,\hat e_{m}\},
$$
 be the local frame for the tangent
bundle  $T\partial M$ and $\hat x=(\hat x^i)=(\hat x^2,\dots,\hat x^{m})$, 
be the local coordinates on $\partial M$.  Let $r$ be the
normal geodesic distance  to $\partial M$,  and 
$$
\hat
N=\partial_r\big|_{\partial M}
$$
 be the inward pointing unit normal to
$\partial M$.   Let $V$ be a (smooth) vector bundle over the manifold $M$,
$\nabla$ be a connection on $V$ and $\hat\nabla$ be the induced  connection on
the boundary. Further let $Q$ be a smooth endomorphism of $V$ and $F$ be a
Laplace type operator 
$$
F=-g^{\mu\nu}\nabla_\mu\nabla_\nu+Q.
$$

Let $W=V\big|_{\partial M}$ be the
restriction of the vector bundle $V$ to the boundary $\partial M$. We define
the boundary data map  $\psi:\ C^\infty(V)\to C^\infty(W\oplus W)$ by   \bes
\psi(\varphi)=\left(\matrix{\varphi|_{\partial M}\cr
\nabla_N\varphi|_{\partial M}\cr}\right)\,. 
\ees
Let $\Pi$ be an orthogonal Hermitian projector acting on $W$  and  
$\Gamma$ be an anti-Hermitian $\End(W)$-valued vector field $\Gamma$ on the
boundary that satisfies the conditions 
\bea
(\Gamma^i)^*&=&-\Gamma^i,\nonumber\\[10pt]
\Pi\Gamma^i&=&\Gamma^i\Pi=0.\nonumber
\eea
 Let, further, $S$ be a smooth
Hermitian endomorphism of the bundle $W$ orthogonal to $\Pi$, i.e.  
\bea
S^* &=& S \nonumber\\[10pt]
\Pi S &=& S\Pi=0. \nonumber
\eea
We use these to define a first-order self-adjoint
tangential differential operator $\Lambda:\ C^\infty(W)\to C^\infty (W)$ by
\bes
\Lambda=(\II-\Pi)\left\{
{1\over 2}(\Gamma^i\hat\nabla_i+\hat\nabla_i\Gamma^i)
+S\right\}(\II-\Pi)\,.
\ees

We study the oblique boundary-value problem for the Laplace type operator
$F$. 
The oblique boundary conditions now read 
\bes
B\psi(\varphi)=0\,,
\ees
where $B: C^\infty(W\oplus W)\to C^\infty(W\oplus W)$ is the boundary
operator defined by
\be
B=\left(\matrix{\Pi&0\cr
\Lambda & \II -\Pi\cr}\right)\,.
\label{5ms}
\ee
This is equivalent to the following boundary conditions
\bea
\Pi \varphi \Big|_{\partial M} &=& 0\,,\nonumber\\[10pt]
(\II-\Pi)\nabla_N\varphi\Big|_{\partial M}
+\Lambda \varphi\Big|_{\partial M} &=& 0 .
\eea
The boundary operator $B$ (\ref{5ms})
incorporates all standard types of boundary conditions. Indeed, by choosing
$\Pi=\II$ and $\Lambda=0$ one gets the Dirichlet boundary conditions, by
choosing $\Pi=0$,  $\Lambda=\II$ one gets the Neumann boundary conditions.
More generally, the choice $\Gamma=0$  corresponds to the mixed boundary
conditions.

Integration by parts shows that the Laplace-type operator $F$  endowed with
oblique  boundary conditions is symmetric, i.e.  
$$
(\varphi,F\psi)  =
(F\varphi,\psi).
$$
However, it is not necessarily elliptic.   To be elliptic
the boundary-value problem $(F,B)$ has to satisfy two conditions. First of
all, the leading symbol of the operator $F$  should be elliptic in the
interior of $M$.  Second, the so-called strong ellipticity condition should be
satisfied.  This question was studied in  \cite{avramidi99a} where it has
been shown that the oblique boundary-value problem for a Laplace type operator
is strongly elliptic with respect to the cone $\CC\setminus \RR_+$ if and only
if for any nonvanishing cotangent vector $\zeta$ on the boundary the
endomorphism $|\zeta| \II-i\Gamma\cdot\zeta$ is positive-definite, i.e.
$|\zeta|\II-i\Gamma\cdot\zeta>0$.   A sufficient condition for strong
ellipticity is: 
$$
|\zeta|^2\II+(\Gamma\cdot\zeta)^2>0.
$$


The heat kernel $U(t|x,y)$ is now defined by the heat equation
\bes
(\partial_t+F)U(t)=0\,,
\ees
the initial condition 
\bes
U(0^+|x,y)=\delta(x,y)\,,
\ees
and the boundary
conditions 
\be
B\psi[U(t|x,y)]=0\,.
\ee
Hereafter the boundary data map (as well as the boundary operator)  acts on
the first argument of the heat kernel.
 
The heat kernel of a  smooth boundary-value problem on a manifold with
boundary has the following asymptotic expansion as $t\to 0^+$ 
\cite{gilkey95}
\be 
\Tr_{L^2}\exp(-tF)\sim
\sum\limits_{k=0}^\infty t^{(k-m)/2}A_{k}\,.  
\label{17}
\ee
In the case of manifolds without boundary  only even order terms
were present (see \ref{71-5/01}).  Now, in contrast, all $A_k$ are non-zero.
They have the following general form: 
\bea
A_{2k} &=& \int\limits_M a^{(0)}_{2k}
+\int\limits_{\partial M}a^{(1)}_{2k},
\nonumber\\[10pt]
A_{2k+1} &=& \int\limits_{\partial M}a^{(1)}_{2k+1}\,.
\nonumber
\eea
Hereafter the integration over the boundary is defined with the help of the
usual Riemannian volume element $d\vol_{\hat g}$ on $\partial M$ with the help
of the induced metric $\hat g$. 

Here $a^{(0)}_{k}$ are the (local) interior heat-kernel coefficients and
$a^{(1)}_{k}$ are the boundary ones.  The local interior coefficients
$a^{(0)}_k$ are determined by the same local invariants  as in the manifolds
without boundary, i.e. by the HMDS  coefficients $\tr_V a^{\rm diag}_k$, and
therefore, do not depend on the boundary conditions. The boundary coefficients
$a^{(1)}_{k}$ are far more complicated because in addition to the geometry of
the manifold $M$ they depend essentially on the geometry of the  boundary
$\partial M$ and on the boundary conditions. For Laplace-type operators they
are known  for the usual boundary conditions (Dirichlet,  Neumann, or mixed
version of them) up to $a^{(1)}_{5}$  \cite{branson90a,branson99}. For oblique
boundary conditions including tangential derivatives some coefficients were
recently computed in \cite{avramidi99a,avramidi99b,dowker97,dowker98}. In this
lecture we will  evaluate the coefficient $A_{1}$, following our recent work
\cite{avramidi99a}.


Let us fix a positive number $\delta>0$. We split the whole manifold in a
disjoint union of two different parts:  
$$
M=M^{\rm int}\cup M^{\rm bnd},
$$
where $M^{\rm bnd}$ is a narrow geodesic strip near the boundary $\partial M$
of the width $\delta$ and $M^{\rm int}$ is the interior of the manifold
$M$ (without the thin strip), i.e. $M^{\rm int}=M\setminus M^{\rm bnd}$.

We will construct the parametrix on $M$ by using different approximations
in $M^{\rm bnd}$ and $M^{\rm int}$. Strictly speaking, to glue them together
in a smooth way  one should use `smooth characteristic functions' of different
domains (partition of unity) and carry out all necessary estimates. What one
has to control is the order of the remainder terms in the limit $t\to 0$ and
their dependence on $\delta$. Since our task here is not to prove the form of
the asymptotic expansion, which is known, but rather to compute
explicitly the coefficients of the asymptotic expansion, we will not worry
about such subtle details. We will compute the asymptotic expansion as $t\to
0$ in each domain and then take the limit $\delta\to 0$.

We will use different local coordinates in different domains. In $M^{\rm
int}$ we can, for example, choose normal coordinates centered at a fixed
point $x_0$. In fact, this is not necessary---we can use a manifestly
covariant technique described in lecture 2. In $M^{\rm bnd}$ we choose the
local coordinates as follows.  By using the geodesic flow we get the local
frame 
$$
\{N,e_i\}
$$
for the tangent bundle $TM$ and the local coordinates
$x=(r,\hat x)$  on $M^{\rm bnd}$, which identifies $M^{\rm bnd}$  with
$\partial M\times (0,\delta)$.


The construction of the parametrix in the interior $U^{\rm int}(t|x,y)$ goes
along the same lines as for manifolds without boundary described in the
previous lectures. The idea is always to separate the basic case (when
the coefficients of the operator $F$ are frozen at a fixed point   $x_0$). In
the case of manifolds without boundary the basic case  is, in fact,
zero-dimensional, i.e.  algebraic. The interior parametrix is defined by the
heat equation (\ref{he-5/01}), the initial condition (\ref{init-5/01}) and by
an asymptotic condition at infinity (instead of the boundary conditions). This
means that effectively one introduces a small expansion parameter 
$\varepsilon$ reflecting the fact that the points $x$ and $y$  are close to
each other and the parameter $t$ is small. This can be done by fixing a point
$x_0=x'$ in $M^{\rm int}$, choosing the normal  coordinates at this point (with
$g_{\mu\nu}(x')=\delta_{\mu\nu}$), scaling like in (\ref{scal-5/01})
and expanding in a power series in $\varepsilon$. This construction is
standard and we do not repeat it here (see lectures 2 and 4). Locally, at any
point in $M^{\rm int}$, the resulting interior parametrix is given by the
same formulas as the heat kernel for a manifold without boundary, i.e. by the
same formulas as in the lecture 2. For a fixed finite $\delta>0$ the error of
this approximation is exponentially small as $t\to 0$. Thus, the interior
parametrix has the same asymptotic expansion with $t\to 0$ as the heat kernel
for a manifold without boundary. In other words, as $t\to 0$
\bea
\int\limits_{M^{\rm int}}\tr_V U^{\rm int}_{\rm diag}(t)
\sim
\sum_{k=0}^\infty t^{k-m/2} (4\pi)^{-m/2}{(-1)^k\over k!}
\int\limits_{M^{\rm int}} \tr_V a^{\rm diag}_k\,,
\label{intpar-5/01}
\eea
where $a^{\rm diag}_k$ are the standard local HMDS coefficients for 
manifolds without boundary computed in lecture 2.
By taking the limit $\delta\to 0$ of this equation we obtain
\bea
a_{2k+1}^{(0)} &=& 0\,,
\label{a2k1-5/01} \\[10pt]
a_{2k}^{(0)} &=& (4\pi)^{-m/2}{(-1)^k\over k!}\tr_V a^{\rm diag}_k\,.
\label{a2k-5/01}
\eea


For an elliptic boundary-value problem  the diagonal of the parametrix
$U_{\rm diag}^{\rm bnd}(t)$ in $M^{\rm bnd}$ has   exponentially small terms,
i.e. of order $\sim\exp(-r^2/t)$, as $t\to 0^{+}$ and $0<r<\delta$. These
terms  behave like distributions near the boundary, and, therefore, the
integrals over $M^{\rm bnd}$, more precisely, the integrals
$$
\lim_{\delta\to 0}\int\limits_{\partial M}
\int\limits_0^{\delta}dr(\dots),
$$
do contribute to the asymptotic expansion
with coefficients being the integrals over $\partial M$. It is this phenomenon
that leads to the boundary terms in the heat kernel coefficients. Thus, such
terms determine the local boundary contributions  $a^{(1)}_{k}$ to the global
heat-kernel coefficients $A_{k}$.

The boundary parametrix $U^{\rm bnd}(t|x,x')$ in $M^{\rm bnd}$ is
constructed as follows.   Now we want to find the fundamental solution of the
heat equation near diagonal, i.e. for $x\to x'$ and for small $t\to 0$ in the
region $M^{\rm bnd}$ close to the boundary, i.e. for  small $r$ and
$r'$, that satisfies the boundary conditions on $\partial M$ and
an asymptotic condition at infinity. We fix a point on the boundary
$x_0\in\partial M$ and choose normal coordinates on $\partial M$  at
this point (with $g_{ij}(0,\hat x_0)=\delta_{ij}$). 

To construct the boundary parametrix, we again scale the coordinates. But now
we include the coordinates $r$ and $r'$ in the scaling
\bea
\hat x &\to& \hat x_0+\varepsilon(\hat x-\hat x_0),\qquad 
\nonumber\\[10pt]
\hat x' &\to& \hat x_0+\varepsilon(\hat x'-\hat x_0)\nonumber\\[10pt]
r &\to& \varepsilon r,\qquad \nonumber\\[10pt] 
r' &\to& \varepsilon r', \qquad\nonumber\\[10pt]
t &\to& \varepsilon^2 t\,.
\label{scalbnd-5/01}
\eea
The corresponding differential operators are scaled by
\bea
\hat\partial &\to& {1\over\varepsilon}\hat\partial, \qquad
\nonumber\\[10pt]
\partial_r &\to& {1\over\varepsilon}\partial_r,\qquad
\nonumber\\[10pt]
\partial_t &\to& {1\over\varepsilon^2}\partial_t\,.
\eea
Then, we expand the scaled operator $F$ in a power 
series in $\varepsilon$, i.e.
\bes
F\sim \sum\limits_{n=0}^\infty\varepsilon^{n-2} F_n,
\ees
where $F_n$ are second-order differential operators with homogeneous symbols.
Next, we expand the scaled boundary operator (with an extra factor
$\varepsilon$ at the operator $\Lambda$) 
\bes
B\sim\sum\limits_{n=0}^\infty\varepsilon^{n} B_{n},
\ees
where $B_{(n)}$ are
first-order tangential operators with homogeneous symbols. At zeroth
order we have 
\bes
F_0=-\partial_r^2-\hat\partial^2, 
\ees
\bes
B_{0}=\left(\matrix{\Pi_0&0\cr \Lambda_0 & \II-\Pi_0\cr}\right),
\ees
where $\Pi_0=\Pi(\hat x')$ and
\bea 
\hat\partial^2 &=& \hat g^{jk}(\hat
x')\hat\partial_{j} \hat\partial_{k},\qquad \nonumber\\[10pt]
\Lambda_0 &=& \Gamma^j(\hat
x')\hat \partial_j\,.  
\eea
Note that
all leading-order operators $F_0$, $B_{0}$ and $\Lambda_0$ have
constant coefficients and, therefore, are very easy to handle.  

The subsequent strategy is rather simple.  We expand the scaled heat kernel
in $\varepsilon$ 
\bes
U^{\rm bnd}\sim\sum_{n=0}^\infty
\varepsilon^{2-m+n}U^{\rm bnd}_{n},
\ees
and substitute into the scaled version of the heat equation and the
boundary condition.  Then, by equating  the like powers in
$\varepsilon$ one gets an infinite set of recursive differential equations 
for $U_{n}$
\bes
(\partial_t+F_0)U^{\rm bnd}_{k}
=-\sum\limits_{n=1}^{k}F_nU^{\rm bnd}_{k-n},
\ees
with the boundary conditions
\bes
B_0\psi[U^{\rm bnd}_{k}]
=-\sum\limits_{n=1}^{k}B_n\psi[U^{\rm bnd}_{k-n}],
\ees
and the asymptotic condition at infinity
\be
\lim_{r\to\infty} U^{\rm bnd}_{k}(t|r,\hat x;r',\hat x')=0\,.
\label{ac-5/01}
\ee

In other words, we decompose the parametrix into the homogeneous parts with
respect to $(\hat x-\hat x_0)$, $(\hat x'-\hat x_0)$, $r$, $r'$ and $t$.
By using this homogeneity we obtain finally the asymptotic expansion of
the diagonal of the boundary parametrix 
\bes
U^{\rm bnd}_{\rm diag}(t)
\sim
\sum_{k=0}^\infty t^{(k-m)/2}
U^{\rm bnd}_{k}\left(1\Big|{r\over\sqrt{t}},\hat x;
{r\over\sqrt{t}},\hat x\right)\,.
\ees
Now we have to integrate the diagonal $U^{\rm bnd}_{\rm
diag}$ over $M^{\rm bnd}$, expand it in an asymptotic series as $t\to 0$, and
then take the limit $\delta\to 0$.  One should stress that the
volume element  $d\vol(x)=\sqrt{|g|}dx$ should also be scaled, i.e. 
\bea
d\vol(\varepsilon r,\hat x)
\sim
d\vol(0,\hat x)\cdot\sum_{k=0}^\infty \varepsilon^k {r^k\over k!}g_k(\hat
x),
\nonumber  
\eea
where
\bes
g_k(\hat x)=
{\partial^k\over\partial r^k}\left[{d\vol(r,\hat x)\over d\vol(0,\hat
x)}\right]\Bigg|_{r=0}\,.
\ees
The coefficients $g_k$  will contribute directly to the coefficients of
the asymptotic expansion. 

We have
\bea
\int\limits_{\rm M^{\rm bnd}} \tr_V U^{\rm bnd}_{\rm
diag}(t)
&\sim&
\sum_{k=0}^\infty t^{(k-m)/2}
\int\limits_{\rm \partial M}\int\limits_0^\delta dr\,
\nonumber\\[10pt]
&&\times
\left[{d\vol(r,\hat x)\over d\vol(0,\hat x)}\right]
\tr_V 
U^{\rm bnd}_{k}\left(1\Big|{r\over\sqrt{t}},\hat x;
{r\over\sqrt{t}},\hat x\right)\,.
\nonumber
\eea
Since as $\delta\to 0$ the volume of $M^{\rm bnd}$ vanishes, i.e.
$$
\lim_{\delta\to 0}\vol(M^{\rm bnd})=0,
$$
the contribution of all regular terms
will vanish in the limit $\delta\to 0$. In contrary, the  singular terms,
which behave like distributions near $\partial M$, will give the $\partial M$
contributions to the boundary heat kernel coefficients $a_k^{(1)}$.
By changing the integration variable $r=\sqrt{t}\xi$ the
integral $\int\limits_0^{\delta}dr(\dots)$ becomes
$$
\int\limits_0^{\delta/\sqrt{t}}d\xi\,t^{1/2}(\dots)
$$ and in the limit
$t\to 0$ becomes the improper integral 
$$\int\limits_0^{\infty}d\xi\,t^{1/2}(\dots)
$$
plus an exponentially small
remainder term. Then in the limit $\delta\to 0$ we obtain integrals over
$\partial M$ up to an exponentially small function that we are not
interested in. More precisely, as the result we get the coefficients
$a_k^{(1)}$ in the form  
\bea
a_k^{(1)}
 & = &
\sum_{n=0}^{k-1}{1\over n!}g_n
\lim_{\delta\to 0}\int\limits_0^{\delta/\sqrt{t}}
d\xi\,\xi^n 
\nonumber\\[10pt]
&&
\times
\tr_VU^{\rm bnd}_{k-n-1}(1|\xi,\hat x;\xi,\hat x)\,.
\label{ak1-5/01}
\eea


In this lecture we will find the boundary parametrix of the heat equation
to leading order, i.e.  $U^{\rm bnd}_0$. We fix a point $\hat x'\in \partial M$
on the boundary and the normal coordinates at this point (with $\hat
g_{ik}(\hat x')=\delta_{ik}$), take the tangent space $T\partial M$ and replace
the manifold $M$ by 
$$
M_0\equiv T\partial M \times \RR_{+}.
$$
By
using the explicit form of the zeroth-order operators $F_0$, $B_0$ and
$\Lambda_0$ we obtain the equation 
\be
\left(\partial_t-\partial_{r}^{2}-\hat\partial^2\right)U^{\rm bnd}_0(t|x,y) =0,
\label{39} 
\ee 
and the boundary conditions 
\bea
\Pi_0U^{\rm bnd}_0(t|x,y)\Big|_{r(x)=0} &=&0, 
\label{40} \\[10pt]
(\II-\Pi_0)\left(\partial_r+i\Gamma^j_0\hat\partial_j\right)
U^{\rm bnd}_0(t|x,y)\Big|_{r(x)=0} &=& 0, 
\label{41} 
\eea
where $\Pi_0=\Pi(\hat x'),\
\Gamma^j_0=\Gamma^j(\hat x')$. As usual the differential operators
always act on the first argument of a kernel.  Moreover, for simplicity
of notation, we will denote $\Pi_0$ and $\Gamma_0$ just by $\Pi$ and
$\Gamma^j$ and omit the dependence of all geometric objects on $\hat
x'$.  To leading order this does not cause any misunderstanding.
Furthermore, the heat kernel should be symmetric and vanish at infinity.

By using the Laplace tarnsform in $t$ and Fourier transform in $(\hat x-\hat
y)$ this equation reduces to an ordinary differential equation
of second order in $r$ on $\RR_+$, which can be easily solved taking into
account the boundary conditions at $r=0$ and $r \to \infty$.  Omitting simple
but lengthy calculations we obtain  
\bea
U^{\rm bnd}_0(t|x,y)  &=&  \int\limits_{{\bf R}^{m-1}}
{d\zeta\over (2\pi)^{m-1}}
\int\limits_{w-i\infty}^{w+i\infty} 
{d\lambda\over 2\pi i}\,
\nonumber\\[10pt]
&&\times
e^{-t\lambda +i\zeta\cdot(\hat x-\hat y)}
G(\lambda|\zeta,r(x),r(y)),
\label{44}
\eea
where $w$ is a negative constant and $G$ is the 
leading-order resolvent kernel in momentum representation. It reads
\bea
\lefteqn{G(\lambda|\zeta,u,v)
={1\over 2\sqrt{|\zeta|^2-\lambda}}}
\nonumber\\[5pt]
&&\times
\Bigg\{\exp\left\{-|u-v|\sqrt{|\zeta|^2-\lambda}\right\}
\nonumber\\[5pt]
&&
+\left[\II -2\Pi+2i\Gamma\cdot\zeta
\left(\II\sqrt{|\zeta|^2-\lambda}-i\Gamma\cdot\zeta\right)^{-1}\right]
\nonumber\\[5pt]
&&\times
\exp\left[-(u+v)\sqrt{|\zeta|^2-\lambda}\right]
\Bigg\},
\nonumber
\eea
where ${\rm Re}\,\sqrt{|\zeta|^2-\lambda}>0$.

By changing the integration variables, deforming the contour of integration
and  computing certain Gaussian integrals, we obtain the heat kernel diagonal
\bea
U^{\rm bnd}_{0}(t|r,\hat x,r,\hat x) &=& (4\pi t)^{-m/2}\Bigg\{\II
\nonumber\\[10pt]
&&
+\exp\left(-{r^2\over t}\right) (\II-2\Pi)
+\Phi\left({r\over\sqrt t}\right)\Bigg\},
\nonumber
\label{51}
\eea
where 
\bea
\Phi(z) &=&
-2\int\limits_{{\bf R}^{m-1}} {d\zeta\,\over \pi^{(m-1)/2}}\,
\int\limits_C{d\omega\over\sqrt\pi}\,
\nonumber\\[12pt]
&&\times
\exp\left[-|\zeta|^2-\omega^2+2i\omega z\right]\,
\Gamma\cdot\zeta\,
(\omega\,\II+\Gamma\cdot\zeta)^{-1}\,.
\eea
Here the contour of integration $C$ comes from $-\infty+i\varepsilon$,
encircles the point $\omega=i|\zeta|$ in the clockwise direction and goes to
$+\infty+i\varepsilon$, with $\varepsilon>0$ a positive infinitesimal
parameter; the contour $C$ does not cross the interval ${\rm Re}\,\omega=0$,
$0<{\rm Im}\,\omega<|\zeta|$, on the imaginary axis and is above all
singularities of the resolvent $G$.


Now by using eq. (\ref{ak1-5/01}) we
obtain the coefficient $a_1^{(1)}$
\bea
a_1^{(1)}
 & = &
(4\pi)^{-m/2}\int\limits_0^\infty  d\xi\,
\tr_V
\left\{e^{-\xi^2}(\II-2\Pi)
+\Phi(\xi)\right\}\,,
\nonumber
\eea
and, finally, by computing the integral over $\xi$ we get
\bea
a^{(1)}_{1}  &=& 
(4\pi)^{-(m-1)/2}{1\over 4}\Bigg\{-\II-2\Pi 
\nonumber\\[10pt]
&&
+2\,\int\limits_{{\bf R}^{m-1}} 
{d\zeta \,\over \pi^{(m-1)/2}}\,
\exp\left[-|\zeta|^2-(\Gamma\cdot\zeta)^2\right]\Bigg\}.
\label{61}
\eea

We now consider two particular cases. First of all, if
the matrices $\Gamma^i$ form an A\-be\-li\-an al\-ge\-bra, i.e.
$$
[\Gamma^i,\Gamma^j]=0,
$$
then the integral
(\ref{61}) is Gaussian and can be easily evaluated explicitly: 
\bes
a^{(1)}_{1}=(4\pi)^{-(m-1)/2}{1\over 4}\left\{-\II-2\Pi
+2(\II+\Gamma^2)^{-1/2}\right\}.
\ees
Another very important case is when the operator $\Lambda$ is a natural Dirac
type operator when the matrices $\Gamma^j$ form a Clifford like algebra
\bes
\Gamma^{i}\Gamma^{j}+\Gamma^{j}\Gamma^{i}
=2\,\hat g^{ij}{1\over (m-1)}\Gamma^{2},
\ees
where $\Gamma^2=\hat g_{ij}\Gamma^i\Gamma^j$.
In this case one obtains
\bea
a^{(1)}_{1} & = & (4\pi)^{-(m-1)/2}{1\over 4}\Bigg\{-\II-2\Pi
\nonumber\\
&&
+2\left(\II+{1\over (m-1)}\Gamma^2\right)^{-(m-1)/2}\Bigg\}.
\nonumber
\eea

Note that the integral (\ref{61}) diverges when the strong ellipticity
condition, $|\zeta|^{2}\II +(\Gamma\cdot\zeta)^2>0$, is violated. This leads
to singularities in the heat kernel coefficients. This is a general feature
of the oblique boundary-value problem.


\section{Heat-Kernel Asymptotics for Non-Smooth Boundary Conditions}

The boundary-value problem studied in the previous lecture was in the smooth
category. A more general (and much more complicated) setting, so called
singular boundary-value problem, arises when either the symbol of the
differential operator or the symbol of  the boundary operator (or the boundary
itself) are not smooth. In this lecture we study a singular boundary-value
problem for a second order partial differential operator of Laplace type when 
the operator itself has smooth coefficients but the boundary operator is not
smooth. The case when the manifold as well as the boundary are smooth, but
the boundary operator jumps from Dirichlet to Neumann on the boundary, is
known in the literature as Zaremba problem. Zaremba problem belongs to a much
wider class of singular boundary-value problems, i.e. manifolds with
singularities (corners, edges, cones etc.). There is a large body of
literature on this subject where the problem is studied from a very abstract
function-analytical point of view (see \cite{gil01} and the
references therein.) However, the study of heat kernel asymptotics of Zaremba
type problems is quite new, and there are only some preliminary results in this
area \cite{dowker00a,dowker00b}. Moreover, compared to
the smooth category the needed machinery is still underdeveloped. In this
lecture we will closely follow our papers \cite{avramidi00c,avramidi01c}.

Let $(M,g)$ be a smooth compact Riemannian manifold of dimension $m$ with
smooth boundary $\partial M$ with a positive-definite Riemannian metric $g$
on $M$ and induced metric $\hat g$ on $\partial M$. In this lecture we
will be dealing with submanifolds of Riemannian manifolds of codimension one
and two. Therefore, we need to fix notation, first of all. With our notation,
Greek indices, $\mu,\nu,\dots$, label the local coordinates on $M$ and range
from 1 through $m$, lower case  Latin indices from the middle of the alphabet,
$i,j,k,l,\dots$, label the local coordinates on $\partial M$ (codimension one
manifold) and range from $2$ through $m$, and lower case Latin indices from
the beginning of the alphabet, $a,b,c,d, \dots$, label the local coordinates
on a codimension two manifold $\Sigma_0\subset \partial M$ that will be
described later and range over $3,\dots,m$. Further,  we will denote by $\hat
g$ the induced metric on the submanifolds (of the codimension one or two). 
We should stress from the beginning that we slightly abuse the notation by
using the same symbols for all submanifolds (of codimension one and two). This
should not cause any misunderstanding since it is always clear from the
context what is meant.

Let $V$ be a vector bundle over the manifold $M$, $\nabla$ be a connection on
$V$ and $\hat\nabla$ be the induced  connection on the boundary. Further let
$Q$ be a smooth endomorphism of $V$ and $F$ be a Laplace type operator
$$
F=-g^{\mu\nu}\nabla_\mu\nabla_\nu+Q.
$$
Let $W=V\big|_{\partial M}$ be the
restriction of the vector bundle $V$ to the boundary $\partial M$ and $N$ be
the inward pointing unit normal to the boundary. We use these to define the
boundary data map $\psi:\ C^\infty(V)\to C^\infty(W\oplus W)$ and the
boundary operator $B:\ C^\infty(W\oplus W)\to C^\infty(W\oplus W)$; the
boundary conditions then are 
$$
B\psi(\varphi)=0.
$$

We always assume the manifold $M$ itself and the coefficients of the operator
$F$ to be smooth in the interior of $M$. If, in addition, the boundary 
$\partial M$ is smooth, and the boundary operator $B$ is a differential
operator  with smooth coefficients, then $(F,B)$ is a smooth local 
boundary-value problem. If the boundary $\partial M$ 
consists of a finite  number of disjoint connected parts, 
$$\partial
M=\cup_{i=1}^n \Sigma_i,
$$
with each $\Sigma_i$ being compact connected
manifold without boundary, 
$$
\partial\Sigma_i=\emptyset
$$  
and
$$
\Sigma_i\cap\Sigma_j= \emptyset
$$
 if $i\ne j$, then one can  impose 
different boundary conditions on different connected  parts of the boundary
$\Sigma_i$. This means that the full  boundary operator decomposes
$$
B=B_1\oplus \cdots \oplus B_n,
$$
with $B_i$ being different boundary
operators acting on different  bundles. If the boundary operators are
smooth (even if different), then such a boundary-value problem is still smooth.

In this lecture we are interested in a different class of boundary conditions.
Namely, we do not assume the boundary operator to be smooth. Instead, we will
study the case when it has discontinuous coefficients.  Such problems
are often also called mixed boundary conditions; to avoid misunderstanding we
will not use this terminology. We impose different boundary conditions
on connected  parts of the boundary, which makes the boundary-value
problem discontinuous. Roughly speaking, one has a decomposition of a
smooth boundary in some parts where different types of the boundary
conditions are imposed, i.e. say Dirichlet or  Neumann. The boundary operator
is then discontinuous at  the intersection of these parts. 
We consider the simplest case when there are just  two components.
We assume that the boundary of the manifold $\partial M$ is decomposed as the
disjoint union  
\be
\partial M=\Sigma_1\cup\Sigma_2\cup\Sigma_0\,,
\ee
where $\Sigma_1$ and $\Sigma_2$ are smooth compact submanifolds  of dimension
$(m-1)$ (codimension $1$ submanifolds), with the same boundary
$$
\Sigma_0=\partial\Sigma_1=\partial\Sigma_2,
$$
that is a smooth compact
submanifold of dimension $(m-2)$ (codimension $2$ submanifold) without
boundary, i.e. 
$$
\partial \Sigma_0=\emptyset.
$$
Let us stress here that when
viewed as sets both $\Sigma_1$ and $\Sigma_2$ are considered to be 
disjoint open sets, i.e. 
$$
\Sigma_1\cap \Sigma_2=\emptyset.
$$

Let $\pi_1$ and $\pi_2$ be the trivial projections of
sections, $\psi$, of a vector bundle $W$ to $\Sigma_i$ defined by 
$$
(\pi_i\psi)(\hat x)=
\psi(\hat x)
$$
if $\hat x\in \Sigma_i$ and 
$$
(\pi_i\psi)(\hat x)=0
$$
if $\hat x
\not\in \Sigma_i$. In other words $\pi_1$ maps smooth sections of the bundle
$W$ to their restriction to $\Sigma_1$, extending them by zero on $\Sigma_2$,
and similarly for $\pi_2$. Let $S\in
C^\infty(\End(W))$ be a smooth Hermitian endomorphism of the vector bundle $W$.

We study the Zaremba type boundary-value problem for the Laplace type operator
$F$. The Zaremba boundary conditions are
\bes
B\psi(\varphi)=0\,,
\ees
where $B: C^\infty(W\oplus W)\to C^\infty(W\oplus W)$ is a Zaremba type
boundary operator defined by
\be 
B=\left(\matrix{\pi_1&0\cr
\pi_2S\pi_2 & \pi_2\cr}\right) .
\label{18ms}
\ee
In other words, we have Dirichlet boundary conditions on $\Sigma_1$ and
Neumann (Robin) boundary conditions on $\Sigma_2$:  
\bea
\varphi\Big|_{\Sigma_1} &=& 0,
\label{19ms}\\[10pt]
(\nabla_N+S)\varphi\Big|_{\Sigma_2} &=& 0. 
\label{20ms}
\eea
In the following, for simplicity, we restrict ourselves to the case $S=0$.
The projectors $\pi_1$ and $\pi_2$ as well as the boundary operator $B$ are
clearly non-smooth (discontinuous) on $\Sigma_0$. 
Note that the boundary conditions are set only on open subsets
$\Sigma_1$ and $\Sigma_2$; the boundary conditions do not say anything about
the boundary data on $\Sigma_0$.

By integrating by parts on $\partial
M$, it is not difficult to check that  the Zaremba type boundary-value problem
$(F,B)$ for a Laplace-type operator with the boundary operator $B$
(\ref{18ms}) is symmetric. One can also show that it is elliptic with respect
to $\CC\setminus\RR_{+}$.

The heat
kernel  is defined by the equation
\be
(\partial_t+F)U(t|x,x')=0,
\ee
with the initial condition
\be
U(0^+|x,x')=\delta(x,x'),
\ee
and the boundary condition 
\be
B\psi[U(t|x,x')]=0.
\ee


Since coefficients of the  boundary operator $B$ are discontinuous on
$\Sigma_0$, Zaremba type boundary-value problem is essentially singular. For
such problems the asymptotic expansion of the trace of the heat kernel has
additional non-trivial logarithmic terms \cite{gil01}, i.e.
\bea
\Tr_{L^2}\exp(-tF_B)
&\sim& 
\sum\limits_{k=0}^{\infty} t^{(k-m)/2}B_{k}
+\log t\sum\limits_{k=0}^\infty t^{k/2}\, H_{k}\,.
\label{32ms}
\eea
Whereas there are some results concerning the coefficients $B_{k}$, almost
nothing is known about the coefficients $H_{k}$. Since the Zaremba problem
is local, or better to say `pseudo-local', all these coefficients have  the
form  
\bea  
B_{2k} &=& \int\limits_M b^{(0)}_{2k}
+\int\limits_{\Sigma_1}b^{(1),1}_{2k} +\int\limits_{\Sigma_2}b^{(1),2}_{2k}
+\int\limits_{\Sigma_0}b^{(2)}_{2k},
\label{38ms}\\[10pt]
B_{2k+1} &=& \int\limits_{\Sigma_1}b^{(1),1}_{2k+1}
+\int\limits_{\Sigma_2}b^{(1),2}_{2k+1}
+\int\limits_{\Sigma_0}b^{(2)}_{2k+1},\\[10pt]
H_{k} &=& \int\limits_{\Sigma_0}h_k\,.
\eea
Here the new feature is the appearance of the integrals over $\Sigma_0$, which
complicates the problem even more, since the coefficients now depend on the
geometry of the imbedding of the codimension $2$ submanifold $\Sigma_0$ in
$M$ that could be pretty complicated, even if smooth.


Let us stress here that we are not going to provide a rigorous construction
of the parametrix with all  the estimates, which, for a singular 
boundary-value problem,  is a task that would require a separate paper.  For
a complete and mathematically rigorous treatment the reader is referred to
\cite{gil01} and references therein.
Here we keep instead to a pragmatic approach and  will describe the
construction of the parametrix that can be used to calculate explicitly
the heat kernel coefficients $B_k$ as well as $H_k$.

First of all, we need to properly describe the geometry of the problem. Let
us fix two small positive numbers $\varepsilon_1,\varepsilon_2>0$. We split
the whole manifold in a disjoint union of four different parts: 
\bea
M &=&
M^{\rm int}\cup M^{\rm bnd}
\nonumber\\[10pt]
&=&
M^{\rm int}\cup M^{\rm bnd}_1\cup M^{\rm bnd}_2 \cup M^{\rm bnd}_0\,.
\nonumber
\eea
Here $M^{\rm bnd}_0$ is defined as the set of points in the narrow strip 
$M^{\rm bnd}$ of the manifold $M$ near the boundary $\partial M$ of the width
$\varepsilon_1$ that are at the same time in a narrow strip of the width
$\varepsilon_2$ near $\Sigma_0$ 
\bea
M^{\rm bnd}_0
&=&
\{x\in M\ |\ {\rm dist}(x,\partial M)<\varepsilon_1,
{\rm dist}(x,\Sigma_0)<\varepsilon_2\}\,.
\nonumber
\eea
Further, $M^{\rm bnd}_1$ is the part of the thin strip $M^{\rm bnd}$ of the
manifold  $M$ (of the width $\varepsilon_1$) near the boundary $\partial M$ 
that is near $\Sigma_1$ but at a finite distance from $\Sigma_0$, i.e.
\bea
M^{\rm bnd}_1
&=&
\{x\in M\ |\ {\rm dist}(x,\Sigma_1)<\varepsilon_1,
{\rm dist}(x,\Sigma_0)>\varepsilon_2\}\,.
\nonumber
\eea
Similarly,
\bea
M^{\rm bnd}_2
&=&
\{x\in M\ |\ {\rm dist}(x,\Sigma_2)<\varepsilon_1,
{\rm dist}(x,\Sigma_0)>\varepsilon_2\}\,.
\nonumber
\eea
Finally, 
$M^{\rm int}$ is the interior of the manifold $M$ without a thin strip at the
boundary $\partial M$, i.e.
\bea
M^{\rm int}
&=&
M\setminus \left(M^{\rm bnd}_1\cup M^{\rm bnd}_2 \cup M^{\rm bnd}_0\right)
\nonumber\\[10pt]
&=&
\{x\in M\ |\ {\rm dist}(x,\partial M)>\varepsilon_1\}\,.
\nonumber
\eea

We will construct the parametrix on $M$ by using different approximations
in different domains. Strictly speaking, to glue them together in a smooth way 
one should use `smooth characteristic functions' of different domains
(partition of unity) and carry out all necessary estimates. What one has to
control is the order of the remainder terms in the limit $t\to 0$ and
their dependence on $\varepsilon_1$ and $\varepsilon_2$.  Since our task here
is not to prove the form of the asymptotic expansion (\ref{32ms}), which is
known, but rather to compute explicitly the coefficients of the asymptotic
expansion, we will not worry about such subtle details. We will compute the
asymptotic expansion as $t\to 0$ in each domain and then take the limit
$\varepsilon_1,\varepsilon_2\to 0$.

We will use different local coordinates in different domains. In $M^{\rm
int}$ we do not fix the local coordinates; our treatment will be manifestly
covariant. 

In $M^{\rm bnd}_1$ we choose the local coordinates as follows. Let 
$$\{\hat
e_i\},
$$
$(i=2,\dots m)$, be the local frame for the tangent bundle 
$T\Sigma_1$ and $\hat x=(\hat x^i)=(\hat x^2,\dots,\hat x^{m})$, 
$(i=2,\dots,m)$, be the local coordinates on $\Sigma_1$.  Let 
$$r={\rm
dist}(x,\Sigma_1)
$$
be the normal distance  to $\Sigma_1$ (
$$
r=0
$$
being the
defining equation of $\Sigma_1$), and 
$$
\hat N=\partial_r\big|_{\Sigma_1}
$$ be
the inward pointing unit normal to $\Sigma_1$.  Then by using the geodesic
flow we get the local frame 
$$
\{N,e_i\}
$$ for the tangent bundle $TM$ and the
local coordinates $x=(r,\hat x)$  on $M_1^{\rm bnd}$. The coordinate $r$
ranges from $0$ to $\varepsilon_1$, 
$$
0\le r\le\varepsilon_1.
$$
The local
coordinates in $M_2^{\rm bnd}$ are chosen similarly.

Finally, in $M^{\rm bnd}_0$ we choose the local coordinates as follows.
Let 
$$
\{\hat e_a(\hat x)\},
$$
$(a=3,\dots,m)$, be a local frame for
the  tangent bundle $T\Sigma_0$ and let $\hat x=(\hat x^a)=(\hat x^3,\dots,\hat
x^{m})$ be the local coordinates on $\Sigma_0$. To avoid misunderstanding we
should stress here that now we use the same notation $\hat x$ to denote
coordinates on $\Sigma_0$ (not on the whole of $\partial M$).  Let
$
\dist_{\partial M}(x,\Sigma_0)
$
be the distance from a point $x$ on
$\partial M$ to $\Sigma_0$ along the boundary $\partial M$. Then define
$$
y=+\dist_{\partial M}(x,\Sigma_0)>0
$$
 if $x\in \Sigma_1$ and
$$
y=-\dist_{\partial M}(x,\Sigma_0)<0
$$
 if $x\in \Sigma_2$. In other
words, 
$$
y=0
$$
 on $\Sigma_0$ (
$$
r=y=0
$$
 being the defining equations of
$\Sigma_0$), $y>0$ on $\Sigma_1$ and $y<0$ on $\Sigma_2$. Let 
$$
\hat n(\hat
x)=\partial_y\big|_{\Sigma_0}
$$
 be the unit normal to $\Sigma_0$
pointing inside $\Sigma_1$. Then by using the tangential geodesic flow
along the boundary (that is normal to $\Sigma_0$) we first get the local
orthonormal frame 
$$
\{n(y,\hat x),e_a(y,\hat x)\}
$$
 for the tangent bundle
$T\partial M$. Further, let the unit normal vector field to the boundary $\hat
N(y,\hat x)$ be given. Then by using the normal geodesic flow
to the boundary we get the local frame 
$$
\{N(r,y,\hat x),n(r,y,\hat
x),e_a(r,y,\hat x)\}
$$
 for the tangent bundle $TM$ and local coordinates
$(r,y,\hat x)$ on $M_0^{\rm bnd}$. 
The ranges of the coordinates $r$ and $y$ are:
$$
0\le r\le\varepsilon_1
$$
 and 
$$
-\varepsilon_2\le y\le\varepsilon_2.
$$
Finally, we introduce the polar coordinates 
\bes
r=\rho\cos\theta, \qquad
y=\rho\sin\theta\,.
\ees
To cover the whole $M^{\rm bnd}_0$  the angle $\theta$ ranges from $-\pi/2$
to $\pi/2$ and $\rho$ ranges from $0$ to some $\varepsilon_3$ (depending on
$\varepsilon_1$ and $\varepsilon_2$), 
$$
0\le\rho\le\varepsilon_3.
$$


The construction of the interior parametrix goes along the same lines as for
manifolds without boundary (see lectures 2 and 4). For a finite
$\varepsilon_1>0$ the diagonal of the heat kernel has the same asymptotic
expansion as for manifolds without boundary. Therefore, by integrating
over the interior part $M^{\rm int}$ and taking the limit $\varepsilon_1\to
0$ we find that the local interior coefficients $b_{k}^{(0)}$ are the same as
for manifolds without boundary in the smooth case, i.e. $a_k^{(0)}$, given by
(\ref{a2k1-5/01})-(\ref{a2k-5/01}).


The Dirichlet parametrix $U^{\rm bnd, (1)}(t)$ in $M^{\rm bnd}_1$ and
Neumann parametrix $U^{\rm bnd, (2)}(t)$ in $M^{\rm bnd}_2$ are constructed
along the same lines as the parametrix of a smooth boundary-value problem
described in lecture 5. For finite $\varepsilon_1, \varepsilon_2>0$ the
diagonal of the parametrix has the same kind of asymptotic expansion as $t\to
0$ with coefficients being homogeneous functions of $r/\sqrt{t}$
\bes
U^{\rm bnd,(i)}_{\rm diag}(t)
\sim
\sum_{k=0}^\infty t^{{(k-m)\over 2}}
U^{\rm bnd,(i)}_{k}\left(1\Big|{r\over\sqrt{t}},\hat x;
{r\over\sqrt{t}},\hat x\right). 
\ees
After integrating the diagonal over $M^{\rm bnd}_i$ and taking the limit
${\varepsilon_1,\varepsilon_2\to 0}$ the contribution of all regular terms will
vanish. The  singular terms, which behave like distributions near $\Sigma_i$,
will give the $\Sigma_i$ contributions to the boundary heat kernel
coefficients $b_k^{(i)}$. 
As the result we get the coefficients $b_k^{(1),1}$ in the form 
\bea 
b_k^{(1),i}
&=&
\sum_{n=0}^{k-1}{1\over n!}g_n\lim_{\varepsilon_1\to 0}
\int\limits_0^{\varepsilon_1/\sqrt{t}}  d\xi\,\xi^n
\nonumber\\[10pt]
&&\times
\tr_VU^{\rm bnd,(i)}_{k-n-1}(1|\xi,\hat x;\xi,\hat x)\,.
\eea
These are the standard boundary heat kernel coefficients for smooth Dirichlet
and Neumann boundary conditions. They are listed for example in
\cite{branson90a,branson99} up to $k=4$. The first two have the form
\bea
b_0^{(1),1}&=&b_0^{(1),2}=0,
\nonumber\\[10pt]
b_1^{(1),1}&=&-b_1^{(1),2}=-(4\pi)^{-(m-1)/2}\,\dim V\,{1\over 4},
\nonumber\\[10pt]
b_2^{(1),1}&=&b_2^{(1),2}= (4\pi)^{-m/2}\,\dim V\, {1\over 3}\, K\,,
\eea
where $K$ is the trace of the extrinsic curvature (second fundamental form) of
the boundary.


The most complicated (and the most interesting) is the case of the mixed
parametrix in $M^{\rm bnd}_0$ since here the basic problem with frozen
coefficients on $\Sigma_0$ is two-dimensional. More precisely, in
$M^{\rm bnd}_0$ the basic problem is on the half-plane.   Since the origin is
a singular point, we will work in polar coordinates introduced above. One can
still use the scaling device described above. Since now we are working in the
vicinity of the submanifold $\Sigma_0$, the coordinate $\rho$ should also be
scaled, i.e.
\bea
\hat x &\to& \hat x_0+\varepsilon(\hat x-\hat x_0),\qquad \nonumber\\[10pt] 
r&\to& \varepsilon r,\qquad\nonumber\\[10pt]
\rho &\to& \varepsilon \rho,\nonumber\\[10pt]
t &\to& \varepsilon^2 t\,,\nonumber
\eea
and similarly for the coordinates $\hat x'$, $r'$ and $\rho'$.
Then one needs to expand in $\varepsilon$ and develop a perturbation theory,
which gives a recursion system that determines all coefficients $B_k$ and
$H_k$. The order at which the $\log t$ terms show up depends on
the dimension of the manifold.

We will restrict ourselves to the zeroth order of this perturbation theory
only. This is enough for computation of the coefficient $B_1$. 
So, we are going to solve only the basic problem for operators with frozen
coefficients at a point $x_0$ on $\Sigma_0$. We choose normal coordinates
on $\Sigma_0$ at this point (with $g_{ab}(0,\theta,\hat x_0)=\delta_{ab}$). 
Then the zeroth order operator $F_0$ has the form  
\be 
F_0=-\partial_\rho^2 - {1\over \rho}\partial_\rho  -{1\over \rho^2}
\partial_\theta^2-\hat\partial^2  
\ee
and the zeroth order normal is
\bea
N_0\Big|_{\Sigma_1}
&=&-{1\over \rho}\partial_\theta\Big|_{\theta={\pi\over 2}}\,,\qquad
\nonumber\\[10pt]
N_0\Big|_{\Sigma_2}
&=& {1\over \rho}\partial_\theta\Big|_{\theta=-{\pi\over 2}}\,.
\eea
Now the boundary operator is discontinuous, and there  is a singularity
at the origin $\rho=0$.

By separating the `free' semiclassical factor due to $\Sigma_0$ we get the
ansatz
\bea
U^{\rm bnd,(0)}_0(t|\rho,\theta,\hat x;\rho',\theta',\hat x')
&=&(4\pi t)^{-(m-2)/2}
\nonumber\\[10pt]
&&\times\exp\left(-{|\hat x-\hat x'|^2\over 4t}\right) 
\Psi(t|\rho,\theta;r',\theta')\,,
\eea
where
$\Psi(t|\rho,\theta;\rho',\theta')$ is a
two-dimensional heat kernel  determined by the equa\-tion   
\bes
\left(\partial_t -\partial_\rho^2-{1\over\rho}\partial_\rho
-{1\over\rho^2}\partial_{\theta}^2\right)
\Psi(t|\rho,\theta;\rho',\theta') = 0 \,,
\ees
the initial condition
\bes
\Psi(0^+|\rho,\theta;\rho',\theta') = {1\over \rho }
\delta(\rho-\rho')\delta(\theta-\theta')\,, 
\ees
the boundary conditions
\bea
\Psi(t|\rho,\theta;\rho',\theta')\Big|_{\theta={\pi\over 2}} &=& 0,
\nonumber\\[10pt]
\partial_\theta
\Psi(t|\rho,\theta;\rho',\theta')\Big|_{\theta=-{\pi\over
2}} &=& 0\,, \nonumber
\eea
and 
the asymptotic condition at infinity
\bes
\lim_{\rho\to\infty}\Psi(t|\rho,\theta;\rho',\theta')=0\,. 
\ees
Clearly, the heat kernel is also symmetric 
$$
\Psi(t|\rho,\theta;\rho',\theta') = \Psi(t|\rho',\theta';\rho,\theta).
$$
This problem can be solved by separating variables, employing the Hankel
transform in the radial coordinate and evaluating a certain spectral series of
Bessel functions. As a result, we obtain the mixed leading parametrix in
$M^{\rm bnd}_0$ 
\bea
U^{\rm bnd,(0)}_0(t|\rho,\theta;\rho',\theta')
&=&
L(t|\rho,\theta,\hat x;\rho',\theta',\hat x')
\nonumber\\[10pt]
&&
+L(t|\rho,\theta,\hat x;\rho',-\theta'-\pi,\hat x')
\nonumber
\eea
where
\bea
L(t|\rho,\theta,\hat x;\rho',\theta',\hat x')
&=&(4\pi t)^{-m/2}
\nonumber\\[10pt]
&&
\times
\exp\left\{-{|\hat x-\hat x'|^2+\rho^2+\rho'^2-2\rho\rho'
\cos(\theta-\theta')\over 4t}\right\} 
\nonumber\\[10pt]
&&\times
\erf\left\{\sqrt{\rho\rho'\over t}
\cos\left({\theta-\theta'\over 2}\right)\right\}\,,
\nonumber
\eea
with 
$$
\erf(z)={2\over\sqrt{\pi}}\int\limits_0^z du\,e^{-u^2}
$$
 being the error function.

The diagonal of the mixed parametrix is easily found to
be  
\bea
U^{\rm bnd,(0)}_{{\rm diag},0}(t) 
&=& (4\pi t)^{-m/2}
\Biggl\{1
\nonumber\\[10pt]
&&-\sign(\theta)
\exp\left(-{\rho^2\cos^2\theta\over t}\right)
-\erfc\left({\rho\over \sqrt t}\right) 
\nonumber\\[10pt]
&&
+\sign(\theta) \exp\left(-{\rho^2\cos^2\theta\over t}\right)
\erfc\left({\rho\over \sqrt t
}\left|\sin\theta\right|\right)\Biggr\}\,,
\eea
where $\sign(x)$ is the sign function, i.e. $\sign(x)=1$ for $x>0$ and
$\sign(x)=-1$ for $x<0$, and 
\bea
\erfc(z) &=& 1-\erf(z) \nonumber\\[10pt]
&=& 2\pi^{-1/2}\int_z^\infty du\,e^{-u^2}
\eea
 is the complementary error function. 

Finally, we compute the integral of the diagonal of the mixed parametrix over
$M^{\rm bnd}_0$ and take the limit $\varepsilon_1,\varepsilon_2\to 0$.
We have
\bes
\lim_{\varepsilon_1,\varepsilon_2\to 0}
\int\limits_{M^{\rm bnd}_0} \tr_VU^{\rm bnd,(0)}_{{\rm
diag},0}(t) 
=
\lim_{\varepsilon_3\to 0}\int\limits_{\Sigma_0}
\int\limits_0^{\varepsilon_3} d\rho\, \rho \int\limits_{-\pi/2}^{\pi/2}d\theta
\tr_V U^{\rm bnd,(0)}_{{\rm diag},0}(t),
\ees
for some finite $\varepsilon_3>\sqrt{\varepsilon^2_1+\varepsilon_2^2}>0$.
By computing the integrals and taking the limits we obtain 
\bes
\lim_{\varepsilon_1,\varepsilon_2\to 0}
\int\limits_{M^{\rm bnd}_0} \tr_VU^{\rm bnd,(0)}_{\rm diag,0 }(t)
= -t^{(2-m)/2}\,(4\pi)^{-m/2}{\pi\over 4}\dim V\,\vol(\Sigma_0)\,.
\ees
This gives exactly the coefficient $b_2^{(2)}$ in the heat trace  asymptotic
expansion (\ref{38ms}), i.e.
\be
b_2^{(2)}=-(4\pi)^{-(m-2)/2}\dim V\,{1\over 16}\,.
\ee

\section*{Acknowledgments}

I would like to thank Giampiero Esposito and the other organizers of the 
workshop ``Quantum Gravity and Spectral Geometry'' for their kind invitation
to present these lectures, and for financial support through the Istituto
Italiano per gli Studi Filosofici and the Azienda Autonoma Soggiorno e
Turismo, Napoli.

\section*{Note Added in Proof}

The Zaremba problem considered in the section 6 was studied recently by
Seeley in \cite{seeley01}. It has been shown that the logarithmic terms in
the expansion (\ref{32ms}) are absent, i.e. $H_k=0$ for any $k$, which confirms
the conjecture of \cite{avramidi00c}. Seeley has also shown that 
the correct setting of the Zaremba problem involves an additional boundary
condition along the singular set $\Sigma_0$. The general solution as well as
the heat kernel asymptotics do depend on this additional condition. Our
solution corresponds to the choice of most regular eigenfunctions close to
$\Sigma_0$. Other solutions contain integrable singularities near $\Sigma$,
which lead to additional contributions to the heat kernel coefficients.
I am very grateful to Peter Gilkey, Gerd Grubb and Robert Seeley for fruitful
and stimulating discussion of this interesting problem.

\end{document}